\documentclass[sigconf]{aamas} 
\usepackage{balance} % for balancing columns on the final page

\usepackage{pifont}
\usepackage{bm}
\usepackage{bbm}
\usepackage{algorithm}
\usepackage{algpseudocode}
\usepackage{marvosym}

\setcopyright{ifaamas}
\acmConference[ACM conf]{ACM Conference}{2026}
{under review}{TBD}
\copyrightyear{2026}
\acmYear{2026}
\acmDOI{}
\acmPrice{}
\acmISBN{}

\setcopyright{none}

\acmSubmissionID{596}

\title[AAMAS-2026 Formatting Instructions]{%Economically Guided Deep Calibration of Multi-Agent Model for Simulating Real-World Stock Trading
Agent-Based Modelling for Real-World Stock Markets under Behavioral Economic Principles}

\author{
Tianlang He$^{1*}$, Fengming Zhu$^{1*}$, Keyan Lu$^1$, Chang Xu$^2$, Yang Liu$^2$, Weiqing Liu$^2$,
Fangzhen Lin$^1$, S.-H. Gary Chan$^1$, Jiang Bian$^2$\textsuperscript{\Letter}
}
\thanks{$^*$Equal contribution; alphabetical order.} 
\thanks{\textsuperscript{\Letter}Corresponding author.}
\affiliation{
  \institution{$^1$The Hong Kong University of Science and Technology}
  \country{}
}
\affiliation{
  \institution{$^2$Microsoft Research Asia}
  \country{}
}
\email{{theaf, fzhuae}@cse.ust.hk}
\email{jiang.bian@microsoft.com}

\begin{abstract}
The reproduction of realistic dynamics in financial markets is of great significance, as it enhances our understanding of market evolution beyond other physical processes, and facilitates the development and backtesting of investment strategies.
Most existing literature approaches this issue as a time series forecasting problem, which often faces challenges such as 1) overfitting historical data, 2) failing to reconstruct stylized facts, and 3) limiting users' ability to conduct counterfactual analyses.
To address these limitations, we employ agent-based modeling (ABM) for market simulation, where each trader acts as an autonomous agent guided by established behavioral-economic principles.
The parameters of the agent model are subsequently calibrated using deep learning techniques.
Additionally, we align our agent model with publicly available economic indices, such as the Consumer Price Index (CPI), to enhance the explainability of our system's outcomes.
Our experiments demonstrate that the ABM method effectively reproduces market dynamics with a confidence level of 90\%, accurately reflecting well-known stylized facts.
Furthermore, the calibration process proves to be more computationally efficient compared to other existing methods that perform simulation-based inference.
We also present case studies illustrating the correlation between agent parameters and economic indices.

\end{abstract}


\newcommand{\BibTeX}{\rm B\kern-.05em{\sc i\kern-.025em b}\kern-.08em\TeX}

\begin{document}

%%% The following commands remove the headers in your paper. For final 
%%% papers, these will be inserted during the pagination process.

\pagestyle{fancy}
\fancyhead{}

%%% The next command prints the information defined in the preamble.

\maketitle 

%%%%%%%%%%%%%%%%%%%%%%%%%%%%%%%%%%%%%%%%%%%%%%%%%%%%%%%%%%%%%%%%%%%%%%%%

\section{Introduction}

\begin{figure}[ht]
  \centering
  \includegraphics[width=80mm]{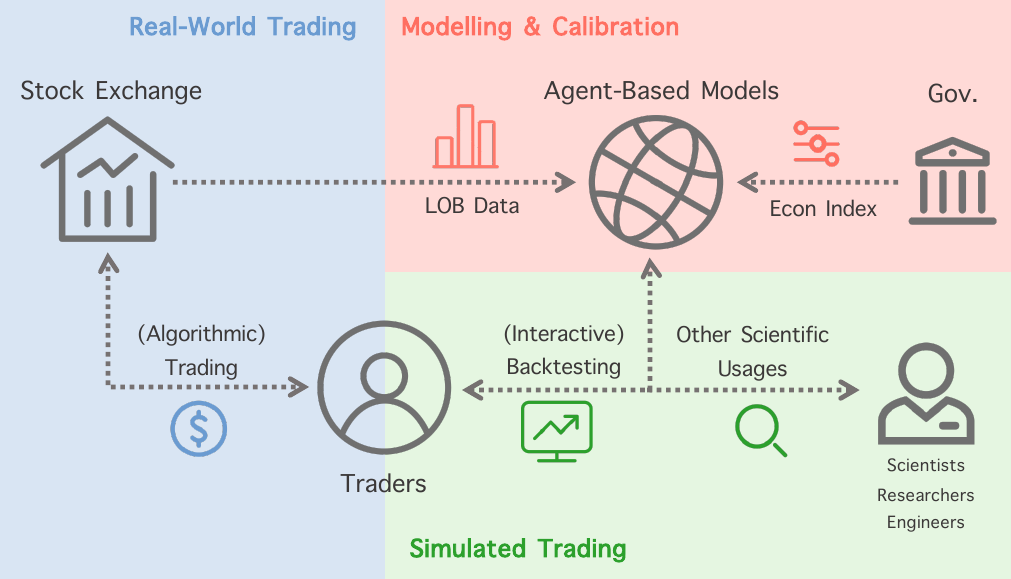}
  \caption{An illustration of the entire ecology. This work focuses on the part of \textit{agent-based modelling} and \textit{calibration}.}
  \label{fig:system}
  \Description{An illustration of the entire system.}
\end{figure}

%%% Background:
% 1. complexity of financial markets
% 2. what people want to do: quantitative research, portfolio management
% 3. significance of a market simulator

Financial markets, particularly those for stock trading, are notorious for their volatility, partly due to the enormous number of participating traders, some of which are even high-frequency computer programs.
Therefore, an effective simulator that can reproduce real-world market dynamics is of great importance.
Firstly, it reveals unique features that distinguish financial processes from other stochastic processes in the physical world.
Additionally, with the help of such simulators, quantitative researchers can conduct backtesting in an interactive manner to evaluate their strategies.

%%% Current literature:
% 1. time series forecasting - cannot do counterfactual analysis
% 2. agent-based modelling - 
% 3. research gap

There are two mainstream ways of implementing a market simulator.
The first method involves viewing it from the perspective of time series forecasting, which can often fit historical data perfectly using parameterized statistical models; however, it prohibits users from performing any counterfactual analysis, such as studying the consequences of \textit{spoofing}\footnote{Spoofing refers to the practice of submitting large spurious orders to buy or sell some security. Real-world examples include the \textit{2010 Flash Crash} in the US.} at a past moment.
The second method is to resort to agent-based models (ABMs), where each trader is modeled as a proactive agent, and market outcomes result collectively from their strategic trading decisions.
 Under this paradigm, while the reproduction of market dynamics may be compromised since each individual agent is often designed as a relatively simpler economic model, counterfactual analysis can be easily conducted by replacing a group of agents with new ones to be tested, which also enables the generation of out-of-distribution samples.
%\textcolor{red}{and out-of-sample tests}

%%% Our method:
% 1. agent model supported by behavioral economics, price and size
% 2. parameters found by deep learning, i.e., MLE
% 3. further fine-tuned to align with market indices

To benefit from both paradigms while addressing their limitations, we primarily follow the second approach but utilize parameterized agent models whose decision-making is informed by well-known theories from behavioral economics~\cite{chiarella1992dynamics,chiarella2006asset,chiarella2009impact,majewski2020co}. In this framework, the bid (or ask) prices are first derived from the economic models, and then the order sizes are correlated with the prices as an indication of risk awareness~\cite{chiarella2009impact}.
The parameters of the agent model are subsequently determined using deep neural networks (DNNs) to ensure that the output market simulations closely resemble real market behavior.
Specifically, the DNN-based calibration module takes real-world \textit{Limit Order Book} (LOB) data as input and outputs the most likely parameters of the agent model that will produce the most realistic market dynamics.
Additionally, we align the agent models with publicly available economic indices, e.g., the \textit{Consumer Price Index} (CPI), to enhance the explainability of our agent-based modeling approach.
One significant challenge is that, although these economic models are parameterized as neural networks, the trading outcomes, which are collectively and temporally determined by the underlying market mechanism, are not differentiable.
To address this issue, we train a surrogate model that directly maps the parameters of the agent models to \textit{stylized facts}, which are statistical regularities observed in most financial markets, based on their resulting market simulations.
Our method successfully achieves a balance among accuracy (through parameterized DNN models), interpretability (theoretically supported by behavioral economics), and interactivity (within the framework of agent-based modeling).

%\textcolor{red}{emphasize our contribution.}

%%% Paper organization

The rest of this paper is organized as follows.
Section~\ref{sec:related} reviews a few related work.
We introduce some formal notations that are useful for describing financial markets in Section~\ref{sec:pre}.
Our system is detailed in Section~\ref{sec:ours} and evaluated in Section~\ref{sec:eval}.
The paper is concluded in Section~\ref{sec:conclusion} with a discussion on potential future directions.

%\textcolor{red}{
%descriptive, predictive, prescriptive
%\\
%Counterfactual analysis
%\\
%Not purely in closed-form equations, in the middle (behavioral econ + parameter estimation), performance + interpretability 
%\\
%Challenges or issues:
%\\
%1. Non-differentiable
%\\
%Network widely accepted for its accuracy generalizability
%}

%%%%%%%%%%%%%%%%%%%%%%%%%%%%%%%%%%%%%%%%%%%%%%%%%%%%%%%%%%%%%%%%%%%%%%%%

%Main references:
%
%LOB~\cite{gould2013limit}
%
%CDA~\cite{friedman2018double}
%
%SG~\cite{shapley1953stochastic}, DP for POSG~\cite{hansen2004dynamic}
%
%The Chiarella model and its variants~\cite{chiarella1992dynamics,chiarella2006asset,chiarella2009impact,majewski2020co}

\section{Related Work}

%\textcolor{red}{Three aspects: Descriptive, Predictive, Prescriptive.}

\label{sec:related}

We are inspired from the following research areas, while the readers should be aware of the differences therein.

\textit{Financial Time Series Forecasting.}
This area primarily focuses on fitting historical data while also accurately predicting the future.
Extensive work has been conducted utilizing statistical models and neural networks to extract and analyze the temporal features of financial markets, primarily concerning market prices and fundamental values~\cite{bai2020entropic,hou2021stock,hou2022multi,xing2023learning,gulmez2023stock}.
Other research focuses on simulating market dynamics by generating the entire order book, including detailed prices and depths of active orders~\cite{li2020generating,coletta2021towards,shi2022state,coletta2022learning}.
This type of research is crucial, as it has become increasingly common to leverage algorithmic trading techniques~\cite{brogaard2023machine,han2023machine,hsu2021fingat,deng2016deep}. Strategies developed must undergo a rigorous backtesting process not only on historical data but also on synthetic out-of-sample data, where simulators are needed, before being deployed in real markets.

\textit{Agent-Based Modelling for Financial Markets.}
One can also employ agent-based models (ABMs) to investigate financial markets~\cite{liu2021finrl,storchan2021learning, karpe2020multi}, as they provide a natural interactive environment for developing new strategies and enable counterfactual analysis.
Examples include studies on the consequences of spoofing~\cite{wang2017spoofing}, market bubbles~\cite{lebaron1999time}, and high-frequency trading~\cite{li2016agent,wah2017latency}, as well as defensive regulations related to real-world incidents, e.g., the 2010 Flash Crash~\cite{paddrik2012agent}.
When real-world data is provided, it is often necessary to first \textit{calibrate} a given simulator, e.g., ABIDES~\cite{byrd2019abides}, which can be achieved via methods like Bayesian inference~\cite{bai2022efficient} or deep learning~\cite{stillman2023deep}.
However, most studies have overlooked the correlation between order prices and sizes.
The literature has also witnessed simulators in other economic fields, e.g., auctions for online advertising~\cite{su2024auctionnet} and  macroeconomic domains like taxation~\cite{mi2024taxai,mi2025econgym}.

%optimization methods: genetic~\cite{mathew2012genetic,alhijawi2023genetic}, ant colony~\cite{blum2005ant,dorigo2006ant}, surrogate-based optoptimization~\cite{queipo2005surrogate,han2012surrogate}

\textit{Strategic Decision-Making.}
In addition to descriptive and predictive studies, prescriptive studies are also significant, as they devise decision-theoretic strategies for individual traders in the market.
The spectrum of such work ranges from the simplest ones, known as the zero intelligence strategy~\cite{bak1997price,maslov2000simple}, to more complext procedures deliberately designed based on iterative observations and belief updates, such as HBL and its variants~\cite{gjerstad1998price,gjerstad2007competitive,tesauro2002strategic,tesauro2001high}.
Recently, a unified framework for long-term strategic decision-making schemes has been proposed and implemented~\cite{zhu2025single}, which may also demonstrate potential in financial applications.
Our work is more fundamental, providing a simulation environment where strategies can be learned or refined with feedbacks.

%%%%%%%%%%%%%%%%%%%%%%%%%%%%%%%%%%%%%%%%%%%%%%%%%%%%%%%%%%%%%%%%%%%%%%%%

\section{Preliminaries}

More than half of the financial markets nowadays operate on \textit{Limit Order Books} (LOBs)~\cite{gould2013limit} under a mechanism termed as the \textit{Continuous Double Auction} (CDA)~\cite{friedman2018double}.
In this section, we first introduce the necessary notations for LOBs and CDA, mainly following the conventions established in~\cite{gould2013limit}.
We then formalize the entire market dynamics in a game-theoretic framework called \textit{Partially Observable Stochastic Game} (POSG)~\cite{shapley1953stochastic,hansen2004dynamic}, followed by the mathematical definition of the research problem addressed in this paper.

\label{sec:pre}

\subsection{Market Dynamics}

\begin{figure}[ht]
  \centering
  \includegraphics[width=80mm]{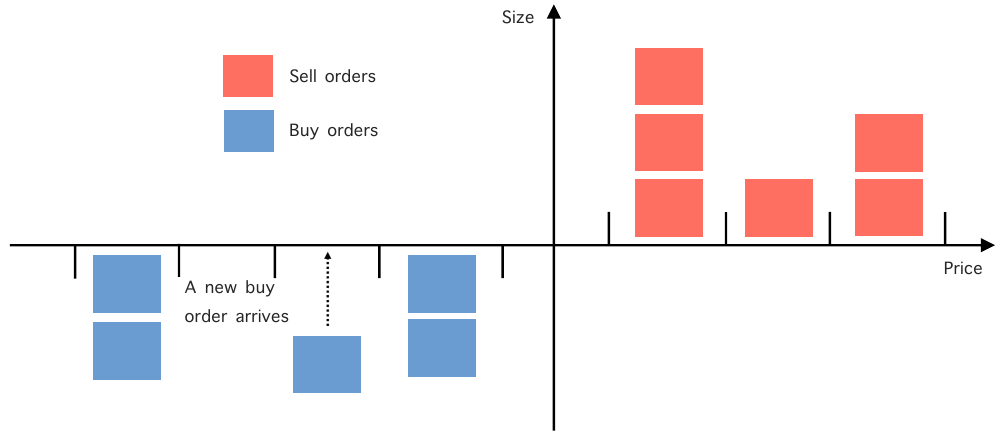}
  \vspace{-1mm}
  \caption{An illustration of the limit order book.
%  \textcolor{red}{modify the font}
  }
  \label{fig:lob}
  \Description{A visualization of the limit order book.}
\end{figure}

A LOB enables every trader to submit either a sell or a buy order with a specified order size at any time.
 Mathematically, a sell (resp. buy) \textit{order} at time $t$ is a tuple $(p_t, v_t)$ with $v_t > 0$ (resp. $v_t < 0$).
 Prices and sizes are not allowed to be any arbitrary quantities, instead, they can only be multiples of pre-specified units, namely the \textit{tick size} $\epsilon$ and the \textit{lot size} $\delta$, respectively.
 The tick size of a LOB is the smallest price increment in a submitted order, and the lot size is the minimum allowable order size, i.e., an order is permissible iff $p_t \in P \triangleq \{k_1\epsilon: k_1 \in \mathbb{Z} \}$ and $v_t \in V \triangleq\{k_2\delta: k_2 \in \mathbb{Z}\}$.
 
 A sell (resp. buy) order that is provisionally not matched with another buy (resp. sell) order is called an \textit{active} order.
Let $\mathcal{L}_t$ denote the set of all active orders at time $t$, $\mathcal{A}_t$ denote the set of all active sell orders, and $\mathcal{B}_t$ denote the set of all active buy orders.
The \textit{mid-price} is defined as the midpoint between the lowest price of active sell order and the highest price of active buy orders, i.e.,
\begin{equation}
m_t = \frac{1}{2}\Big (\min_{(a, x) \in \mathcal{A}_t} a + \max_{(b, y) \in \mathcal{B}_t} b \Big)
\label{eq:midprice}
\end{equation}
The ask-side \textit{depth} for a given price is defined as the total size of all the orders at that sell price, i.e.,
\begin{equation}
n^a_{t}(p) = \sum_{\{(p_t, v_t) \in \mathcal{A}_t: p_t=p\}} v_t,
\end{equation}
similarly for the bid-side depth, denoted as $n^b_{t}(p)$.
For convenience, we refer to $n^a_{t}$ as the set of ask-side depths of all the prices of active sell orders, and similarly for $n^b_{t}$.
%In the rest of this paper, we refer to the \textit{shape} of the order book at time $t$ as the tuple $(\mathcal{A}_t, \mathcal{B}_t, n^a_{t}, n^b_{t})$.
In this paper, we do not consider short sales.

Besides a data structure like a LOB, a market mechanism is needed, i.e., a trade-matching algorithm.
Here we adopt the CDA, as it is prevalent in most financial and commodity markets and is relatively easy to implement.
Basically, when a new sell order is submitted to the market, if there exist any active buy orders currently stored in the LOB with prices higher than that of the new order, then the ones with the highest ``priorities'' will be matched with this new sell order and will be removed from the LOB.
Priorities are given to those active buy orders with the highest prices, and ties are broken by choosing the ones with the earliest submission time.\footnote{Termed as the \textit{price-time} priority mechanism. Other examples include the \textit{pro-rata} and the \textit{price-size} mechanisms.}
Formally, given this newly submitted sell order $(p_t, v_t)$ with $p_t \leq \max_{(b, y) \in \mathcal{B}_t} b$, when such a matching occurs, the highest buy price after the CDA will be $\max(p_t, p_\downarrow)$, where
\begin{equation}
p_\downarrow = \arg\max_{p'} \sum_{p=p'}^{\max_{(b, y) \in \mathcal{B}_t} b} |n^b_{t}(p)| > v_t
%\textcolor{red}{[problematic]}
\end{equation}
We assume the priority ranking is subsumed in the above matching.
If the new sell order does not match any active buy orders, or if there are residual lots of this order after the CDA at this moment, it will be added to the LOB as an active sell order.
The logic is similarly reversed for newly arrived buy orders

Please note that a market order typically does not require a specified price and will be immediately matched with active orders (with the price determined accordingly).
Thus, a market order can be considered as a special case of a limit order.

\subsection{The POSG Framework}
Now we model the markets with the aforementioned features in a game-theoretic framework, termed \textit{Partially Observable Stochastic Game} (POSG), to facilitate the definition of our research problem.
A POSG is defined as a 6-tuple $\langle \mathcal{N}, \mathcal{S}, \mathcal{O}, \mathcal{A}, T, \Omega\rangle$, where
\begin{itemize}
	\item $\mathcal{N}$ is the finite set of agents, where an agent is a trader in the stock market.
	\item $\mathcal{S}$ is the set of all possible states.
	A state $S\in \mathcal{S}$ contains the current information of the LOB, i.e. the state at time $t$ is $\mathcal{L}_t$.
	\item $\mathcal{O} = \prod_{i\in \mathcal{N}} \mathcal{O}_i$, where each $\mathcal{O}_i$ is the set of observations that describe the part of the information exposed to agent $i$.
	In real-world scenarios, this is done by assigning different levels of ``privileges'' to traders.
	Unprivileged traders may have access only to a temporally delayed LOB, an incomplete LOB, or merely some statistics of the current LOB, among other options. In our design, we assume that every trading agent can only access the mid-price at each time, i.e., $m_t$.
	\item $\mathcal{A} = \prod_{i\in \mathcal{N}}\mathcal{A}_i$, where each $\mathcal{A}_i$ is the set of legal actions. We assume $A_1 \equiv \cdots \equiv \mathcal{A}_{|\mathcal{N}|} \triangleq P \times V$, i.e., all agents have the identical set of actions, and a $nil$ action is allowed (i.e., doing nothing by submitting a zero-size order at any price), but order cancellation is not taken into consideration.
	\item $T: \mathcal{S} \times \mathcal{A} \mapsto \mathcal{S}$ is the transition function. The transition of states are the evolution process of $\mathcal{L}_t$ under the CDA rules.
	% any randomness: e.g., fat-finger error?
	% a RCLL process mathematically.
	\item $\Omega = \prod_{i\in \mathcal{N}}\Omega_i$ is the observation function in the form of $\Omega_i: \mathcal{S}_i \mapsto \mathcal{O}_i$. As mentioned, we assume that each agent has access only to the mid-price of the LOB, then each $\Omega_i$ is identical and is exactly given by Equation~(\ref{eq:midprice}). 
\end{itemize}
We omit the components of \textit{rewards} and \textit{discount factors} as they are irrelevant to this work.
Also note that an agent may leave the market temporarily and later rejoin, which is equivalent to continuously performing the \textit{nil} action during her absence.

A policy of agent $i$ is denoted as $\pi_i: (\mathcal{O}_i\times \mathcal{A}_i)^*\times \mathcal{O}_i \mapsto \Delta(\mathcal{A}_i)$, mapping all possible histories to the agent's next (potentially randomized) actions.
We denote $\pi_{\theta_i}$ to refer to a parameterized model with agent $i$'s parameter $\theta_i$, and $\bm{\pi}_\theta$ for a joint parameterized policy.
Given a joint policy $\bm{\pi}_\theta$, a trajectory can be generated, denoted as
\begin{equation}
\big \{(S_t, o_t, a_t) \big \}_t \triangleq
\Big \{\mathcal{L}_t, m_t, \big \{(p_{t,i}, v_{t,i})\big \}_i \Big \}_t
\label{eq:trajectory}
\end{equation}
where each $S_t \in \mathcal{S}$, $o_t \in \mathcal{O}$, and  $a_t \in \mathcal{A}$.
Note that we slightly reload the notation $(p_{t,i}, v_{t,i})$ to indicate that the order is submitted by agent $i$.

The problem that we are faced with is presented in a reversed manner.
That is, given a dataset $\mathcal{D}$ consisting of multiple trajectories with only sequences of states (i.e., actions are missing), we aim to find the most likely parameter $\theta$ such that the generated state sequences closely resemble the given ones.
Formally, the \textit{objective} is to compute
\begin{equation}
\begin{split}
& \arg\min_{\hat\theta}
\mathbb{E}_{
\{\hat{S}_t\}_t \sim \bm{\pi}_{\hat\theta},
\{S_t\}_t \sim \mathcal{D}
}
\Big[ d\big( \{\hat{S}_t\}_t, \{S_t\}_t \big) \Big]
\\=
& \arg\min_{\hat\theta}
\mathbb{E}_{
\{\hat{\mathcal{L}}_t\}_t \sim \bm{\pi}_{\hat\theta},
\{\mathcal{L}_t\}_t \sim \mathcal{D}
}
\Big[ d\big( \{\hat{\mathcal{L}}_t\}_t, \{\mathcal{L}_t\}_t \big) \Big]	
\end{split}
\label{eq:obj}
\end{equation}
where $d$ is a distance measure.
When the context is clear, we use $\mathcal{L}$ to denote $\{\mathcal{L}_t\}_t$ for convenience.
This task is also known as the problem of \textit{model calibration}.
Note that any economic problem that fits within the framework of POSG and the solution formulation of Equation~(\ref{eq:obj}) can be solved by the method that we will propose in the remainder of this paper.
Therefore, calibrating a stock market simulator can be considered as an application of the above principle.

%%%%%%%%%%%%%%%%%%%%%%%%%%%%%%%%%%%%%%%%%%%%%%%%%%%%%%%%%%%%%%%%%%%%%%%%

\section{Our system}

\label{sec:ours}

\begin{figure*}[ht]
  \centering
  \includegraphics[width=0.95\linewidth]{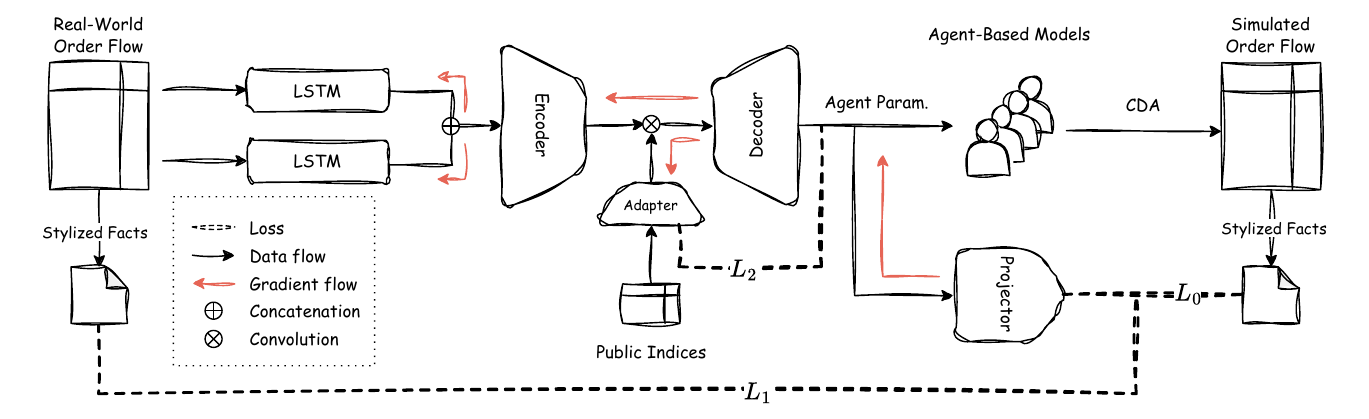}
  \vspace{-2mm}
  \caption{The pipeline of calibrating our agent-based models.
%  \textcolor{red}{three losses and data/gradient flows. Find better icons for order flow}
%  \textcolor{teal}{agent model and then CDA}
}
  \label{fig:pipeline}
  \Description{The entire pipeline of our system.}
\vspace{-2mm}
\end{figure*}

In this section, we present our system by
(1)~introducing the behavioral economic theories that support the agent-based models,
(2)~demonstrating how to implement neural modules to estimate the parameters of those economic models, and
(3)~showing a method for aligning the pre-trained economic model with economic indices to enhance explainability.

\subsection{The Behavioral-Economic Agent Model}

As previously mentioned, the action of each trading agent  consists of both the quoted price and the order size.
The fundamental logic is that an agent first assesses her expected price of the asset based on historical market data, and then selects an order size that depends on the gap between the current price and her expected price, as well as the variation in past prices.
This second step is often overlooked in the literature; however, we argue that it is essential because real-world traders tend to be risk-averse, and incorporating this price-size correlation helps reproduce more realistic market dynamics.

We adopt a well-known paradigm~\cite{chiarella2006asset,majewski2020co} that partitions the agents into three classes, namely \textit{fundamentalists}, \textit{chartists}, and \textit{noise traders}.
Each has its own way of predicting the expected price, and the eventual expectation of an agent is a linear combination of those three predictions, denoted as $\hat{P}^F_t$, $\hat{P}^C_t$, and $\hat{P}^N_t$, respectively.
A \textit{fundamentalist} agrees with the fundamental value of the asset, which is an exogenously assigned signal, given as
$$\hat{P}_t^{F} = \mu_{t},$$ where $\mu_t$ represents the fundamental value, which is typically unobservable.
However, nowadays many financial institutions will publish (or more likely, sell) their own reports with some fundamental analysis.
A \textit{chartist} (also known as a trend-follower) uses a limited period of past history
to predict the future, given as
$$\hat{P}_t^{C} = g(m_{t-\tau}, \cdots, m_{t-1}),$$
where $g$ is a function for regression, and the market prices of the past $\tau$ steps is selected for regression.
A \textit{noise trader} (also known as a zero intelligence trader) is modeled as sampling from a Gaussian distribution, specifically,
$$\hat{P}_t^N \sim \texttt{Normal}(m_{t-1}, \sigma_N)$$
The eventual expected price is a weighted average of the above three predictions, i.e.,
\[
%\hat{P}_t = \frac{\alpha_F\cdot  \hat{P}_t^{F} + \alpha_C\cdot  \hat{P}_t^{C} + \alpha_N\cdot  \hat{P}_t^N}{\alpha_F + \alpha_C + \alpha_N}
\hat{P}_t = \alpha_F\cdot  \hat{P}_t^{F} + \alpha_C\cdot  \hat{P}_t^{C} + (1 - \alpha_F - \alpha_C )\cdot  \hat{P}_t^N
\]

\begin{figure}[h]
  \centering
  \includegraphics[width=60mm]{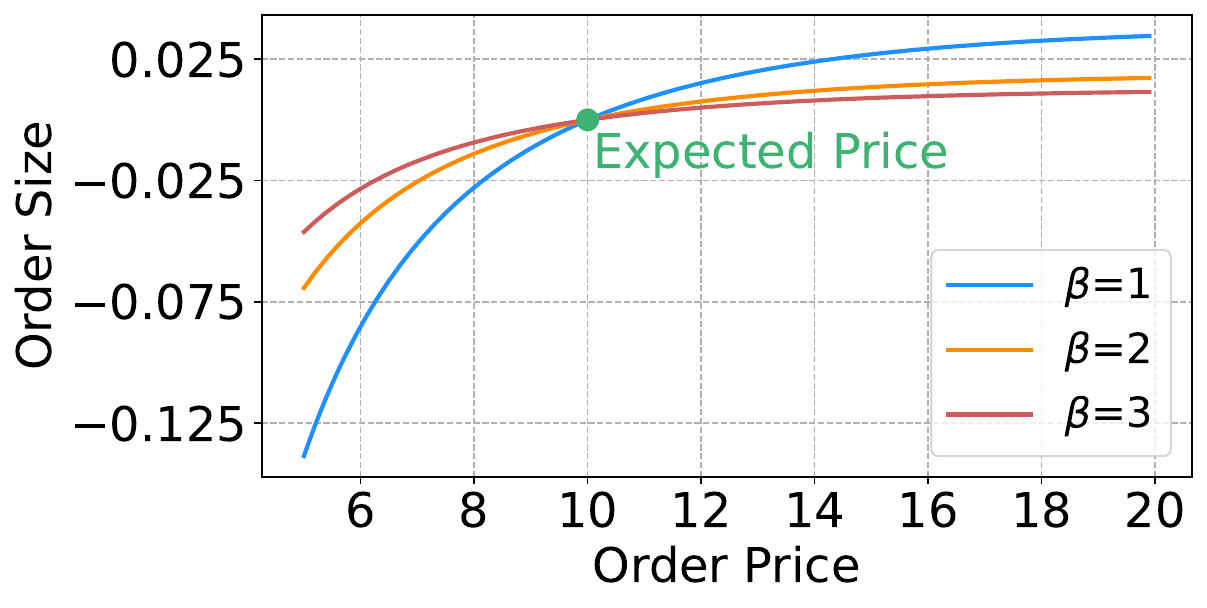}
  \vspace{-1mm}
  \caption{An illustration of how prices and order sizes are correlated under the CARA utility function, with $\hat P_t = 10$, $\Delta m_t = 1$, and sampled $\beta$ varying from $1$ to $3$.}
  \label{fig:cara}
  \Description{xxx}
\end{figure}

Each agent then submits an order with the price $p_t$ sampled from a uniform distribution centered at the expected price, while the size $v_t$ follows from the so-called \textit{Constant Absolute Risk Aversion} (CARA) utility function~\cite{chiarella2009impact},
%\textcolor{red}{unit issue},
\begin{equation}
\begin{split}
& p_t \sim \texttt{Unif}((1 - \eta)\hat{P}_t, (1 + \eta)\hat{P}_t) \\
& \beta \sim \texttt{Normal}(\beta_r, \sigma_r) \\	
& v_t = \frac{\log(p_t / \hat{P}_t)}{\beta \Delta m_t p_t} \\
%& \text{\textcolor{red}{how to ensure prices/sizes are multiples of lot/tick sizes}}
\end{split}
\end{equation}
where $\Delta m_t$ is the variance of the market price up to the current time, and $\beta$ is the degree of risk aversion sampled from a Gaussian distribution.
We provide an illustration in Figure~\ref{fig:cara} showing how CARA works.
Note that the sampled values of $p_t$/$v_t$ will be rounded to the nearest multiples of the tick/lot sizes.
The above economic model applies to all agents, therefore, we omit the agent index. For example, $\hat{P}_t^{F}$ is actually a shorthand for $\hat{P}_{t, i}^{F}$ and similarly for the others.
As there are quite a few parameters involved, to not confuse the readers, we explicitly list the parameters that we will later estimate, and the hyper-parameters that are fixed, i.e.,
\[
\begin{split}
\hat\theta & = \hat\theta_{estimate} \cup \theta_{hyper} \\
\hat\theta_{estimate} & = \{\alpha_F,  \alpha_C, \tau, \sigma_N, \beta_r \} \\
\theta_{hyper} & = \{g, \eta, \sigma_r\} % any others
\end{split}
\]
where $\hat\theta$ refers to the parameters included in Equation~(\ref{eq:obj}).
In our later experiments, we choose
$g(\cdot)$ as linear regression, $\eta = 10\%$, $\sigma_r = 1\% \times \hat P_t$.
For simplicity, we sometimes use ``agent parameters'' as a shorthand of the parameters of agent models.

Intuitively, the heterogeneity among agents arises from these sampling processes,
and market oscillations result from the competition between value and trend, along with the volatility introduced by noise traders~\cite{majewski2020co}.

% each agent uses identical modules and (hyper-)parameters

\subsection{Parameter Estimation by Deep Learning}

We then illustrate mainly the two following aspects:
(1) the network design for parameter estimation,
(2) the surrogate module that bridges the gap of non-differentiability between the parameters of the agent models and the simulated market dynamics.
The main procedure is summarized in Figure~\ref{fig:pipeline} and  Algorithm~\ref{alg:train}.

%\begin{algorithm}[!ht]
%\caption{Training Pipeline \textcolor{red}{conciser?}}
%\label{alg:train}
%\begin{algorithmic}[1]
%\Require A dataset of real-world order flow $\mathcal{L}$
%\Ensure The estimated parameters of the agent models $\hat\theta_{estimate}$
%	\While{not convergence} \Comment Train the Projector
%		\State Sample agent parameters, train $\texttt{Projector}$
%		\Comment Eq.~(\ref{eq:projector1})
%	\EndWhile
%	
%	\While{not convergence} \Comment Train the VAE backbone
%		\State Fix $\texttt{Projector}$, train VAE
%		\Comment Eq.~(\ref{eq:lstm_vae}) and (\ref{eq:projector2})
%	\EndWhile
%	
%	\While{not convergence} \Comment Fine-tune the Adapter
%		\State Fix $\texttt{Projector}$, train VAE and \texttt{Adapter}
%		\Comment Eq.~(\ref{eq:lstm_vae}) and (\ref{eq:adapt})
%	\EndWhile
%\end{algorithmic}
%\end{algorithm}

\begin{algorithm}[!ht]
\caption{Training Pipeline}
\label{alg:train}
\begin{algorithmic}[1]
\Require A dataset of real-world order flow $\mathcal{L}$
\Ensure The estimated parameters of the agent models $\hat\theta_{estimate}$
\State Sample agent parameters, train $\texttt{Projector}$
		\Comment Eq.~(\ref{eq:projector1})
	
\State Fix $\texttt{Projector}$, train VAE
		\Comment Eq.~(\ref{eq:lstm_vae}) and (\ref{eq:projector2})

\State Fix $\texttt{Projector}$, fine-tune VAE and \texttt{Adapter}
		\Comment Eq.~(\ref{eq:lstm_vae}) and (\ref{eq:adapt})

\end{algorithmic}
\end{algorithm}

\subsubsection{The VAE Backbone}
The overall task is to reconstruct market dynamics given the dataset of real-world records, while the reconstruction is done by first estimating the parameters of the agent models and then simulating the interactions of those agents under the CDA.
Thus, a \textit{Variational Auto-Encoder} (VAE)~\cite{kingma2013auto} is a natural choice, as it is commonly used in reconstruction and generation tasks.
With a VAE as the key component, we implement a preprocessing module based on LSTM~\cite{hochreiter1997long} to extract temporal features of the LOB.
For a given trading day, let $t_{max}$ denote the length of the trading records.
Moreover, let $P_t \triangleq \{p\}_{(p, v) \in \mathcal{L}_t}$ denote the vector containing all possible prices of the current active orders, and $V_t \triangleq \{n^a_{t}(p)\}_{(p, v) \in \mathcal{A}_t} \cup \{n^b_{t}(p)\}_{(p, v) \in \mathcal{B}_t}$ denote the vector containing all the depths of the current active orders.
The feedforward computation of the involved networks is done as follows,
\begin{equation}
\begin{split}
h_P & = \texttt{LSTM}_1(P_0, \cdots, P_{t_{max}-1})\\
h_V & = \texttt{LSTM}_2(V_0, \cdots, V_{t_{max}-1})\\
h_{LOB} & = \texttt{concatenate}(h_P, h_V) \\
\mu_0, \sigma_0 & = \texttt{Encoder}(h_{LOB}) \\
z_0 & = \mu_0 +  \rho \cdot \sigma_0,\ \rho \sim \texttt{Normal}(0, 1) \\
\hat{\theta}_{estimate} & = \texttt{Decoder}(z_0)
\end{split}
\label{eq:lstm_vae}
\end{equation}
where the second last line is the so-called \textit{reparameterization} trick, enabling the backpropagation of gradients.

\subsubsection{Tackling the Non-Differentiability}
Recall that the trajectory rollout $\hat{\mathcal{L}}\sim \bm{\pi}_{\hat\theta}$ given the policies of the agents is defined in Equation~(\ref{eq:trajectory}) and (\ref{eq:obj}).
However, although one can calculate the similarity (i.e., reconstruction error) between the generated $\hat{\mathcal{L}}$ with ground truth $\mathcal{L} \sim \mathcal{D}$, the interactions among agents under the rules of CDA is not differentiable, which prohibits gradients from flowing back from the similarity score.
To resolve this issue, we circumvent it by training a surrogate module to directly map $\hat{\theta}_{estimate}$ to certain regularities of its simulated trajectory.
Some of these statistical regularities observed in real-world stock markets are well studied, usually termed as \textit{stylized facts}~\cite{gould2013limit,vyetrenko2020get}.
Four types of stylized facts are taken into consideration,
\begin{enumerate}
	\item \emph{Heavy tails of the return distribution.} The (logarithmic) return of the mid-prices between time $t$ and time $t + \Delta t$ is defined as
	\[
	r_{t, \Delta t} \triangleq \ln (m_{t+\Delta t}) - \ln (m_t).
	\]
	The distribution of returns typically displays a tail heavier than that of a Gaussian distribution, where the heaviness is usually captured by the \textit{kurtosis}\footnote{https://en.wikipedia.org/wiki/Kurtosis} of the return distribution.
	Overall, we calculate three statistics for the collected return distribution: (a) the kurtosis (b) the gain-loss ratio, and (c) the ration of zero returns.
    \item \emph{Volatility clustering.} Many studies on real-world markets have indicated that a large price change tends to be followed by another large price change, partly due to the trend-following behavior of human traders.
    The clustering phenomenon is usually quantified as the autocorrelation of the squared returns. Mathematically, given a return sequence of total length $L$,
    \[
    AutoCorr(l) \triangleq \frac{1}{L - l}
    \sum_{i=1}^{L - l} (r^2_{t_i, \Delta t} - \bar r)(r^2_{t_{i + n}, \Delta t} - \bar r),
    \]
    where $\bar r = \frac{1}{L}\sum_{i=1}^L r^2_{t_i, \Delta t}$.
    Overall, we use $l = 1, 2$ and $3$ as three aggregated statistics.
    \item \emph{Order book depths (volumes).}
    Volumes at the highest bids (resp. the lowest asks) are usually observed to follow Gamma distributions in real-world stock markets~\cite{bouchaud2002statistical}. 
    Therefore, we use the form of Gamma distribution to fit the simulation outcome, and extract the distribution parameter as the aggregated statistic.
    \item \emph{Order sizes.} Real-world order sizes are empirically found to be power-law distributed~\cite{bouchaud2018trades}. Thus, we use the form of the power-law distribution to fit the simulation outcome and extract the distribution parameter as the aggregated statistic.
\end{enumerate}
Other stylized facts include \textit{long memory in order flow}, \textit{time correlation of order flow}, and so on (cf.~\cite{gould2013limit,vyetrenko2020get}).
Upon our experiments, we end up selecting the above four.

The surrogate model takes the estimated parameters of the agent models as input and outputs the predicted stylized facts of the simulated market trajectory generated by these agent models under the CDA.
The role of this surrogate model is to parameterize the interactions of the agents under the given market mechanism.
The feedforward computation of the projector is given as 
\begin{equation}
\begin{split}
\tilde\phi & = \texttt{Projector}(\tilde{\theta}_{estimate}) \\
\tilde\phi^* & = \texttt{CompStylizedFacts}(\tilde{\mathcal{L}}),\ 
\tilde{\mathcal{L}} \sim \pi_{\tilde{\theta}_{estimate} \cup \theta_{hyper}}
\label{eq:projector1}
\end{split}
\end{equation}
where $\tilde{\theta}_{estimate}$ is the uniformly sampled agent parameters.
The projector is supposed to be trained \textit{in advance} by uniformly sampling sufficiently many agent parameters within a certain range,
with the training objective of minimizing the distance between the predicted features and the statistics of the stylized facts of the simulated market dynamics, i.e., minimizing $L_0 = \|\tilde\phi -  \tilde \phi^* \|_2^2$.
Once the projector is trained upfront, it will be frozen and appended to the pipeline after Equation~(\ref{eq:lstm_vae}) to train the VAE backbone, i.e.,
\begin{equation}
\begin{split}
\hat\phi & = \texttt{Projector}_{frozen}(\hat{\theta}_{estimate}) \\
\phi & = \texttt{CompStylizedFacts}(\mathcal{L})
\label{eq:projector2}
\end{split}
\end{equation}
The eventual loss function is designed to guide the VAE backbone so that the estimated agent parameters will reproduce the stylized facts of the real-world market records, i.e., minimizing $L_1 =  \|\hat\phi -  \phi \|_2^2$.
%\textcolor{red}{training detail missing: first train surrogate, and then VAE.}

\subsection{Alignment with Economic Indices}

In addition to the influence of market price fluctuations, it is believed that exogenous signals that published by the government or some financial institutions also affect the investment behavioral of the society, e.g., economics indices like \textit{Consumer Price Index} (CPI), \textit{Producer Price Index} (PPI), \textit{Purchasing Managers' Index} (PMI), and so on.
For instance, CPI is typically calculated monthly or quarterly as a weighted average of consumer expenditures across various categories.
A high CPI may indicate inflation, prompting individuals to shift their investments toward assets with higher fundamental values, which could potentially lead to increased market volatility.
Therefore, we advance our research by not only calibrating agent models to reproduce realistic market dynamics but also investigating how these public indices may influence the parameters of the agent models (and to what extent).

With these indices in hand, we first implement a neural module as an adapter to map them into the latent space of the VAE,
 where we assume that the latent features extracted from these indices are uniformly distributed.
Before decoding into agent parameters, the latent features from public indices are combined with the latent features encoded from the order flow.
Formally, we modify the last few lines in the feedforward computation shown in Equation~(\ref{eq:lstm_vae}) to the following
\begin{equation}
\begin{split}
\cdots & \textit{ (same above) }  \cdots \\
\mu_0, \sigma_0 & = \texttt{Encoder}(h_{LOB}) \\
z_0 & = \mu_0 +  \rho \cdot \sigma_0,\ \rho \sim \texttt{Normal}(0, 1) \\
\mu_1, \sigma_1 & = \texttt{Adapter}(s) \\
z_1 & = \mu_1 + w \cdot \sigma_1,\ w \sim \texttt{Unif}(-k, k) \\
\hat{\theta}_{estimate} & = \texttt{Decoder}(z_0 + z_1) \\
\end{split}
\label{eq:adapt}
\end{equation}
where $s$ denotes the input index, and $k$ is a \textit{consistency range} controlling the extent to which the indices influence the order flow features in the latent space.
There is no ground truth for conducting supervised learning, as the underlying correlation between these indices and the agent parameters is unknown \textit{a priori}.
Nevertheless, an important observation is that under similar market conditions, traders may exhibit similar behavior.
That is, closer indices may indicate similar parameters for the agent model.
Accordingly, one can formulate the following loss function by viewing this problem from the perspective of unsupervised clustering,
%\textcolor{red}{emphasize that agent parameters vary every day}
\begin{equation}
	L_2 = \sum_{x\in \hat\theta_{estimate}} \sum_{i,j}^{|\mathcal{D}|}\frac{\|x_i - x_j\|}{\|s_i - s_j\| + 1},
%	\textcolor{red}{\text{please check the first summation}}
\end{equation}
and the total loss will be
\begin{equation}
L = L_1 + \lambda L_2
\label{eq:final_loss}
\end{equation}
where $\lambda=0.1$ in our later experiments.
Note that the alignment with economic indices occurs after the training of the VAE backbone is completed, and is therefore regarded as a fine-tuning step.
In our experiments, this fine-tuning step typically requires only a very short period of time.

%%%%%%%%%%%%%%%%%%%%%%%%%%%%%%%%%%%%%%%%%%%%%%%%%%%%%%%%%%%%%%%%%%%%%%%%

\section{Empirical Evaluation}

\label{sec:eval}

%Abides
%
%account initialization
%
%Hypo testing, significance level
%
%%%% other details
%
%Data format
%
%Number of agents
%
%
%Advantages over iterative search
%
%Generalization
%
%learning from demonstrations

In this section,
experiments are conducted to showcase the efficiency, effectiveness, and explainability of our proposed ABM system.
\textit{Due to the limited space, we defer some of the detailed training settings to Appendix~\ref{app:train_details}.}

\subsection{Baselines}

Efficient calibration of complex agent-based models using large-scale real-world market data is a relatively new topic in the AI community; therefore, only a limited number of methods are suitable for comparison.
Here, we refer to two baselines methods,
\begin{enumerate}
	\item \textit{Random Search} (RandSearch). It randomly searches for parameters and retains the better ones. Intuitively, it functions similarly to a genetic algorithm that optimizes the objective without utilizing gradients.
	\item \textit{Bayesian Optimization} (BayesOpt)~\cite{bai2022efficient}. This method first requires the assumption of appropriate prior distributions for the parameters and the set-up of a Bayesian network. The posterior distributions are then updated based on the simulation outcomes and the predefined probabilistic dependencies.
\end{enumerate}
Both of the above methods suffer from significant efficiency issues.
RandSearch lacks a good optimization signal, and therefore it normally requires relatively more iterations to run.
BayesOpt relies on the deliberate design of prior/sampling distributions for Bayesian inference (i.e., they often need to be statistical conjugates), rendering the updates even more time-consuming.
In contrast, our method, which employs deep neural networks (DNNs), naturally utilizes gradients to guide optimization in a likelihood maximization framework, eliminating the need for prior assumptions.
Additionally, these two baselines are implemented solely on CPUs, whereas our DNN-based system can take advantage of GPU acceleration.

%\textcolor{red}{Do both of these two baselines only focus on calibration? Does that mean the same economic model is used, i.e. both correlating prices and sizes?}

%\textcolor{red}{Does our DNN-based estimation show better generalization?}

\subsection{Experimental Settings}

\subsubsection{Datasets}

Our study utilizes the following two sets of training data; see Appendix~\ref{app:data_format} for specific data schemas.
\begin{enumerate}
	\item Real-world LOB-level trading data collected from the A-share market over the whole year of 2020 (254 days in total, discretized by milliseconds per day), with the fundamental values taken as the mid-prices every 10 minutes.
	\item Economic indices that are either publicly available\footnote{https://akshare.akfamily.xyz/data/index.html}, including CPI, PPI, and PMI, or easily calculated, such as \textit{Market Trend} and \textit{Market Noise}.
	CPI, PPI, and PMI are recorded monthly.
	Market Trend is calculated as the monthly price change over the \textit{average true range}~\cite{wilder1978new} for that month.
	Market Noise is measured using the monthly \textit{efficiency ratio}~\cite{kaufman2013trading}.
\end{enumerate}

\subsubsection{Metrics} We incorporate three metrics to verify our simulation outcome, including the alignment with those economic indices, 
\begin{enumerate}
%	\item \emph{Reconstruction error}
%	\item \emph{Behavioral variation}
	\item \emph{Mid-price deviation} measures the average difference in mid-price between the generated order flows and the target flows over a trading day, calculated as
    \begin{equation}
        D_P=\frac{1}{T}\sum_{t=1}^T \frac{\left|m_t-\hat{m}_t\right|}{m_t}.  
    \end{equation}
    \item \emph{Discrepancy of order-book shape} quantifies the difference between the predicted and target order books throughout a trading day.
    We measure the discrepancy using the Kolmogorov-Smirnov (KS) statistic, calculated as
    \begin{equation}
        D_V=\frac{1}{T}\sum_{t=1}^T \sup_p |\operatorname{CDF}_{V_t}(p)-\operatorname{CDF}_{\hat{V}_t}(p)|,
    \end{equation}
    where $\operatorname{CDF}(\cdot)$ denotes the cumulative distribution function of the order depth over the prices at each time step.
    In the experiment, we set the confidence level of the KS test to $90\%$, resulting in a critical value of the KS statistic of $0.36$. 
    In other words, $D_V<0.36$ is considered an insignificant difference between the generated and target order flows. 
    \item \emph{Pearson correlation coefficient} (PCC) measures how well each agent parameter correlates with each economic index over the whole year. The PCC of two random variables $X, Y$ is defined as the covariance divided by the product of their respective standard deviations,
\[
PCC(X, Y) = \frac{Cov(X,Y)}{\sigma_X \sigma_Y}
\]
%i.e.,
%\begin{equation}
%    PCC(a_x,s) = \frac{Cov(a_x, s)}{\sigma_{a_x} \sigma_s}. 
%\end{equation} 
\end{enumerate}

%\subsubsection{Evaluation Protocol \textcolor{red}{this whole subsection}}
%
%1. net arch, hyper-parameter setting, hardware device: for projector (surrogate model), vae, adaptor, respectively.
%$\lambda = 0.1$ for Equation~(\ref{eq:final_loss})
%
%2. agent account initialization, scale=500, other LOB/ABIDES settings
%
%3. training/validating protocol: using five-fold cross-validation, testing set, 5 iterations per epoch.
%
%4. seconds per iteration, 20-30 minutes per simulation
%
%5. Anything that deserves to be mentioned about surrogate model training  

\subsection{Overall Quality of Market Simulations}

\begin{figure}[ht]
  \centering
  \includegraphics[width=60mm]{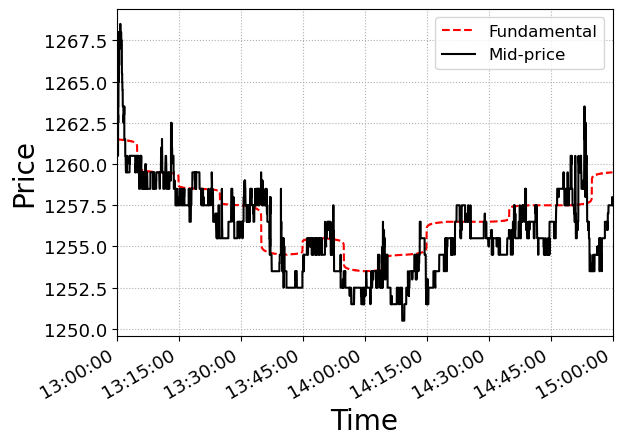}
  \vspace{-2mm}
  \caption{An illustration of the simulated outcome for half of a trading day.}
  \label{fig:simulation}
  \Description{An illustration of the simulated outcome for half of a trading day.}
\end{figure}

We first present a sample output of the simulated trajectory for half a trading day in Figure~\ref{fig:simulation}. 
To enhance readability, we only display the simulated mid-prices and the fundamental values from the dataset, which are recorded every 10 minutes (i.e., the ground truth curve of the mid-prices is omitted).
One can see the simulated mid-prices fluctuate around the fundamental values while also demonstrating some degree of volatility.
\textit{Due to the page limit, we postpone some results on other simulated results to Appendix~\ref{app:stylized}.}

\begin{figure}[ht]
  \centering
  \includegraphics[width=60mm]{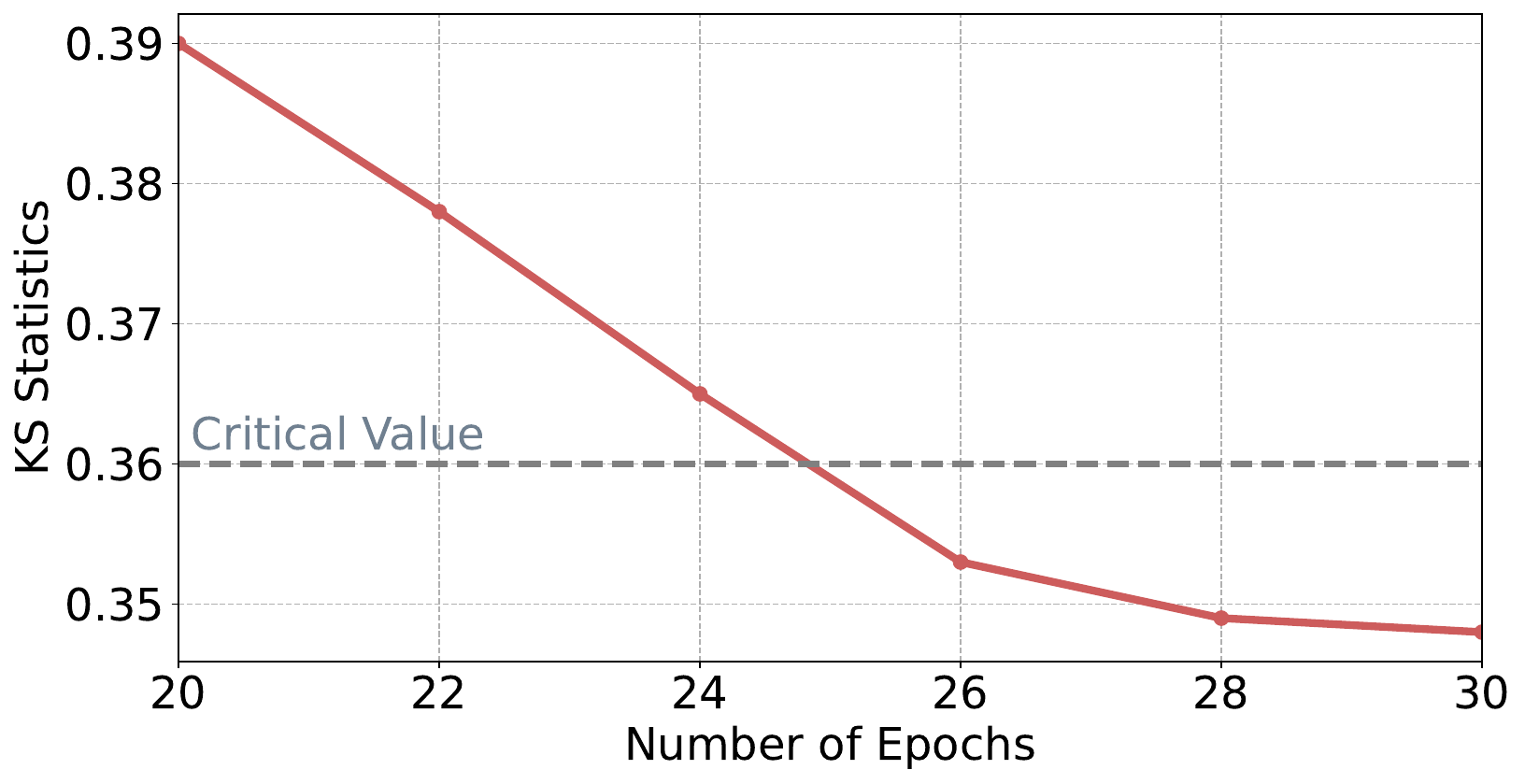}
  \vspace{-2mm}
  \caption{KS statistics on the validation set during training.}
  \label{fig:train_log}
  \Description{KS statistics on the validation set during training.}
\end{figure}

The training of this calibration system is quite efficient.
As shown in Figure~\ref{fig:train_log}, the KS statistic on the validation dataset falls below 0.36 (corresponding to a 90\% confidence level that the simulated trajectory matches the ground truth trajectory) after only 25 training epochs.
We end up with 30 training epochs, achieving a final KS statistic of 0.348.
\textit{Please note that, an epoch consists of only 5 iterations each taking only less than 6 seconds, and therefore, it takes less than 30 seconds to complete a single epoch.}

\begin{figure}[ht]
  \centering
  \includegraphics[width=40mm]{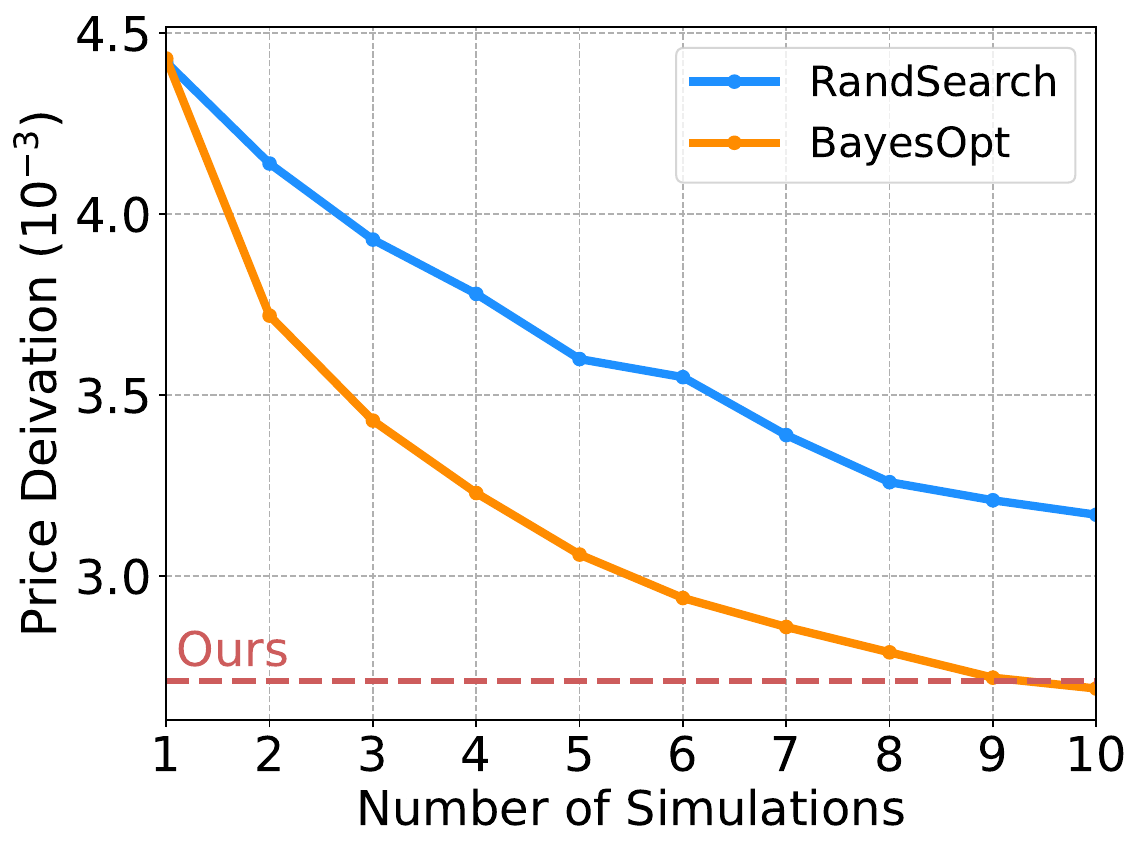}
  \includegraphics[width=40mm]{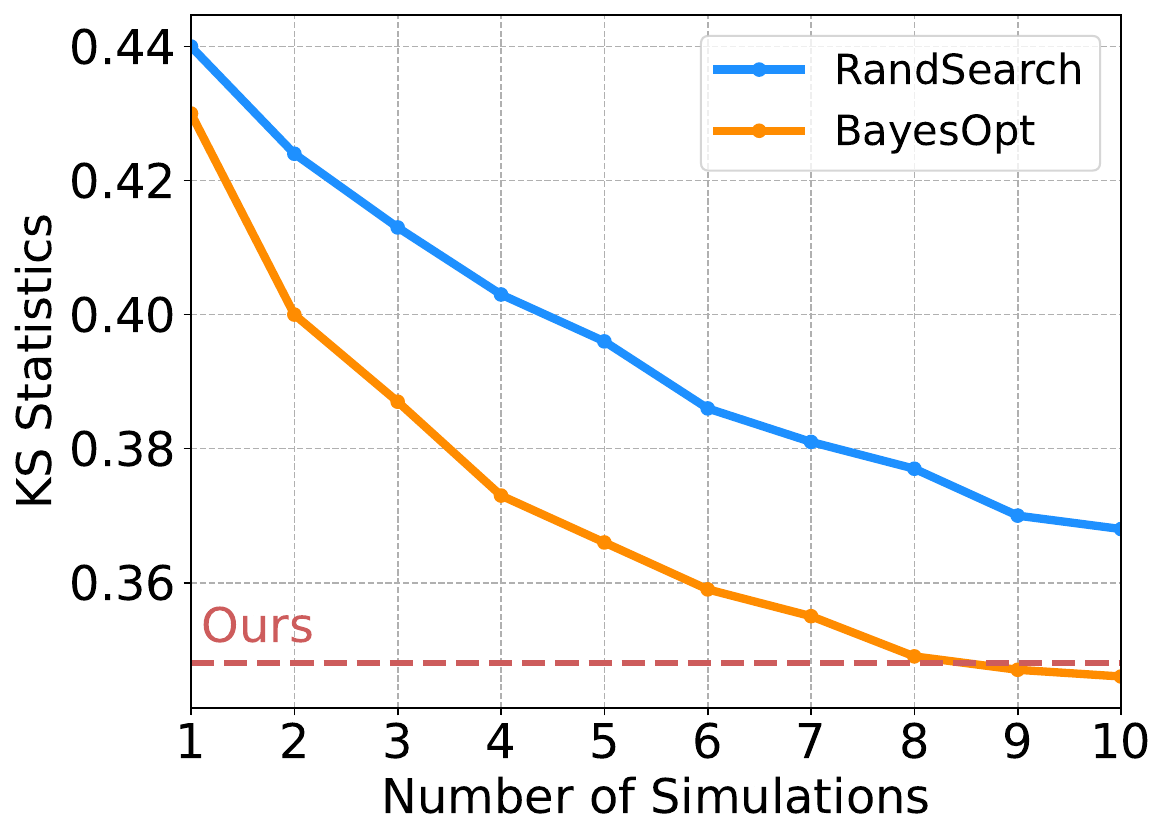}
  \caption{Comparison with the other two calibration methods, in terms of two metrics: price deviation (left), and the KS statistic (right).}
  \label{fig:metric_comparison}
  \Description{Comparison with the other two calibration methods, in terms of two metrics: price deviation (left), and the KS statistic (right).}
\end{figure}

One advantage of a DNN-based ABM is that, once training is complete, the inference time is nearly negligible, given the records from any trading day (or a relatively shorter/longer trading period).
 In contrast, existing work that relies on simulation-based inference must re-collect a sufficient number of simulated samples each time new records from a specific trading period are provided, which is time-consuming, especially when the simulator is not efficient enough.
Evidence is presented in Figure~\ref{fig:metric_comparison}.
\textit{Even for informed search methods like BayesOpT, approximately 10 simulations are required to achieve comparable accuracy to ours, which typically requires around 5 hours to run.}
Note that the collection of simulated samples is generally not parallelized, as parameter updates depend on the outcomes of previous simulations.
One might argue that we also need to collect a set of simulation trajectories to train our surrogate model.
However, readers should be aware that this process can be highly parallelized in our method.
Another perspective on this issue is that methods like RandSearch and BayesOpt incur a constant marginal cost for collecting simulations to calibrate their models on new trading records, while our method conducts this process upfront, resulting in only a fixed cost.

%training/inference time cost: \textcolor{red}{missing now, partly answer by Figure~\ref{fig:train_log}, but could be supplemented by 3 below.}

%training/validation metric over the training phase: Figure~\ref{fig:train_metric_comparison}.
%\textcolor{red}{how to also incorporate the training logging of DeepCal, for an apple-to-apple comparison? E.g., number of iterations/epochs x computational time per iteration/epoch}

\begin{figure}[ht]
  \centering
  \includegraphics[width=40mm]{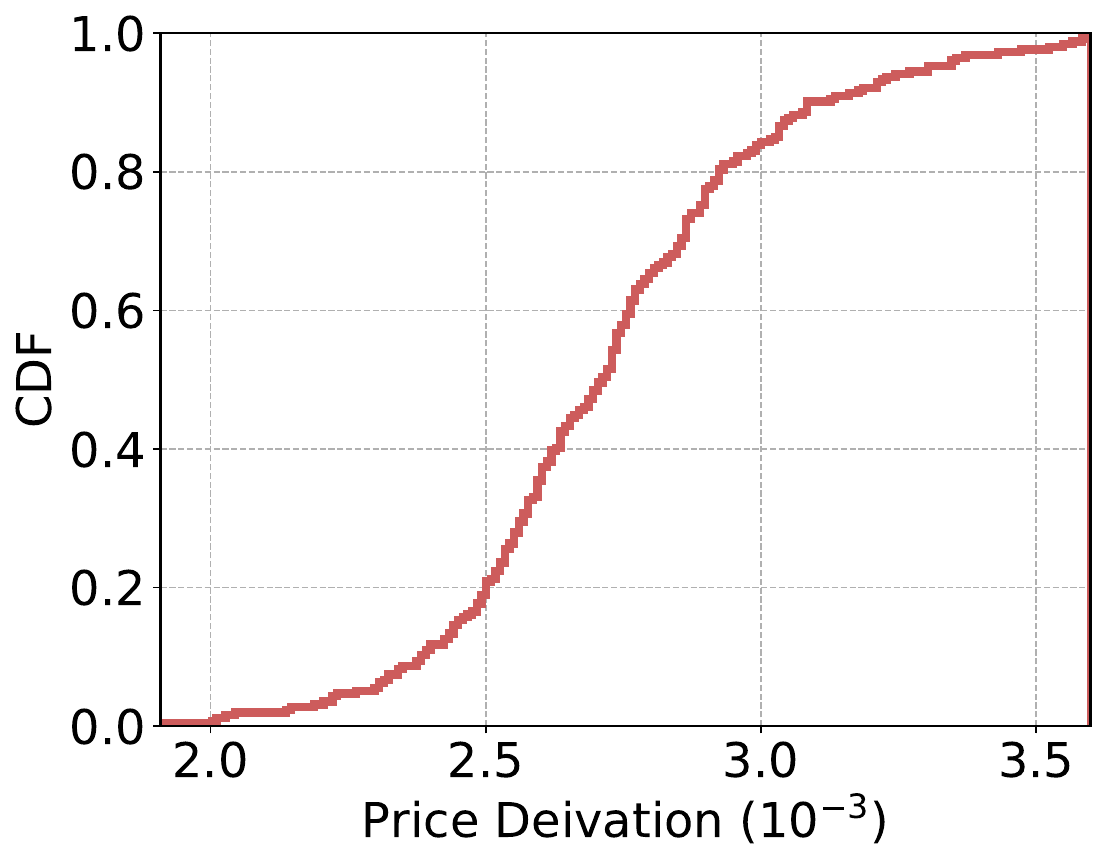}
  \includegraphics[width=40mm]{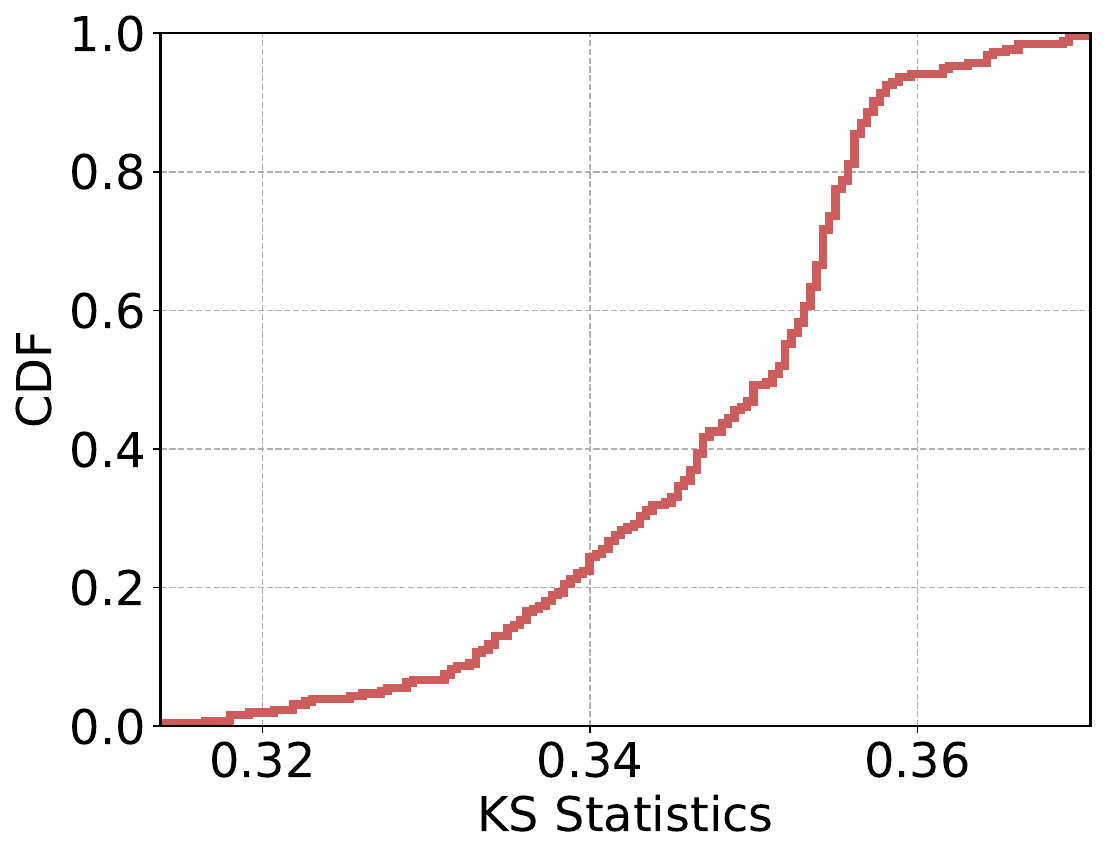}
  \vspace{-2mm}
  \caption{Distributions behind the averages.}
  \label{fig:cdf}
  \Description{xxx}
\end{figure}

Additionally, we present the detailed distributions of the two metrics calculated over the trading days in the test dataset in Figure~\ref{fig:cdf}.
Approximately 80\% of the simulated outcomes are associated with a price deviation of less than $3.0 \times 10^{-3}$, and 90\% of the simulated outcomes are associated with a KS statistic of less than $0.36$.
%\textcolor{red}{distribution over which days, only those from the test set or all the trading days?}

%4. metric CDF comparison: Figure~\ref{fig:cdf}. \textcolor{red}{Can we also include the CDFs for the two baselines? Distribution behind the mean}

\vspace{-2mm}

%\clearpage

\subsection{Case Studies on Alignment with Indices}

This section primarily investigates two aspects:
1) how the alignment with these economic indices affects the reproduction of market dynamics, and  
2) the degree to which the agent parameters are correlated with these indices.
\textit{We first note that this fine-tuning process requires only 6 epochs, five iterations each, to converge.}

\begin{figure}[ht]
  \centering
  \includegraphics[width=40mm]{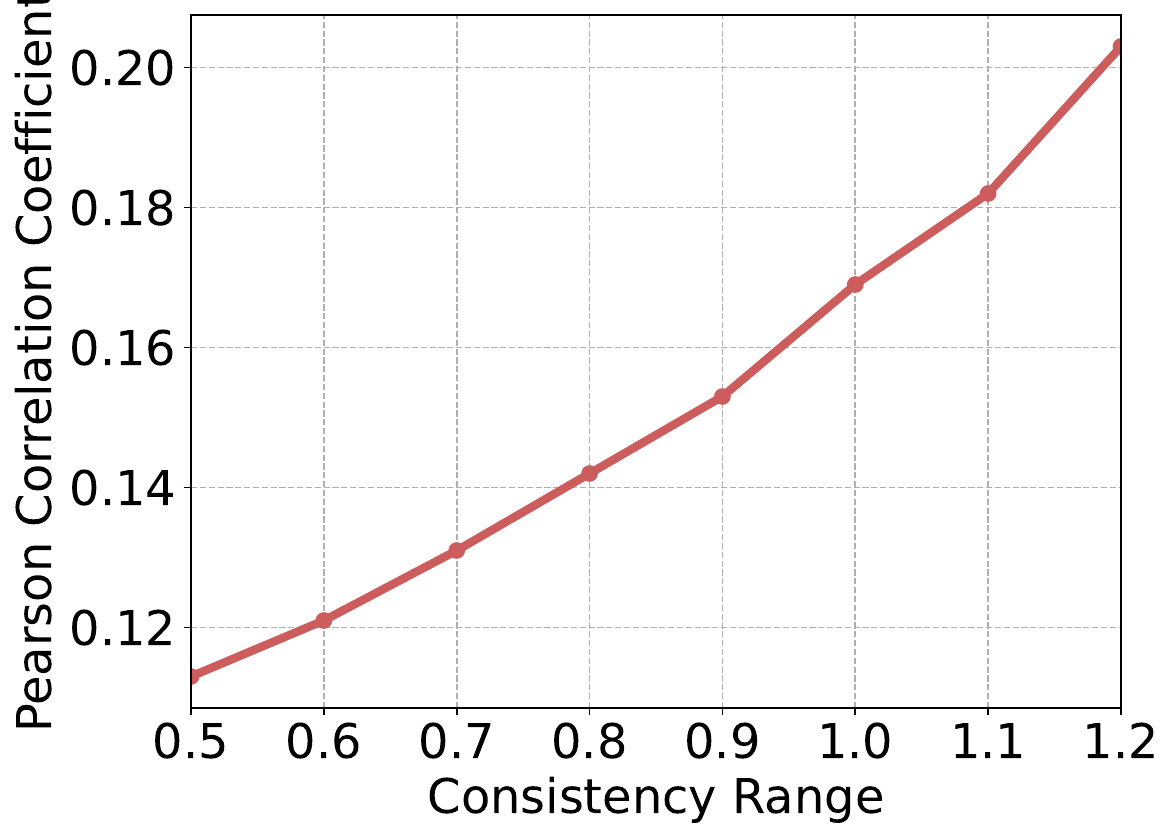}
  \includegraphics[width=40mm]{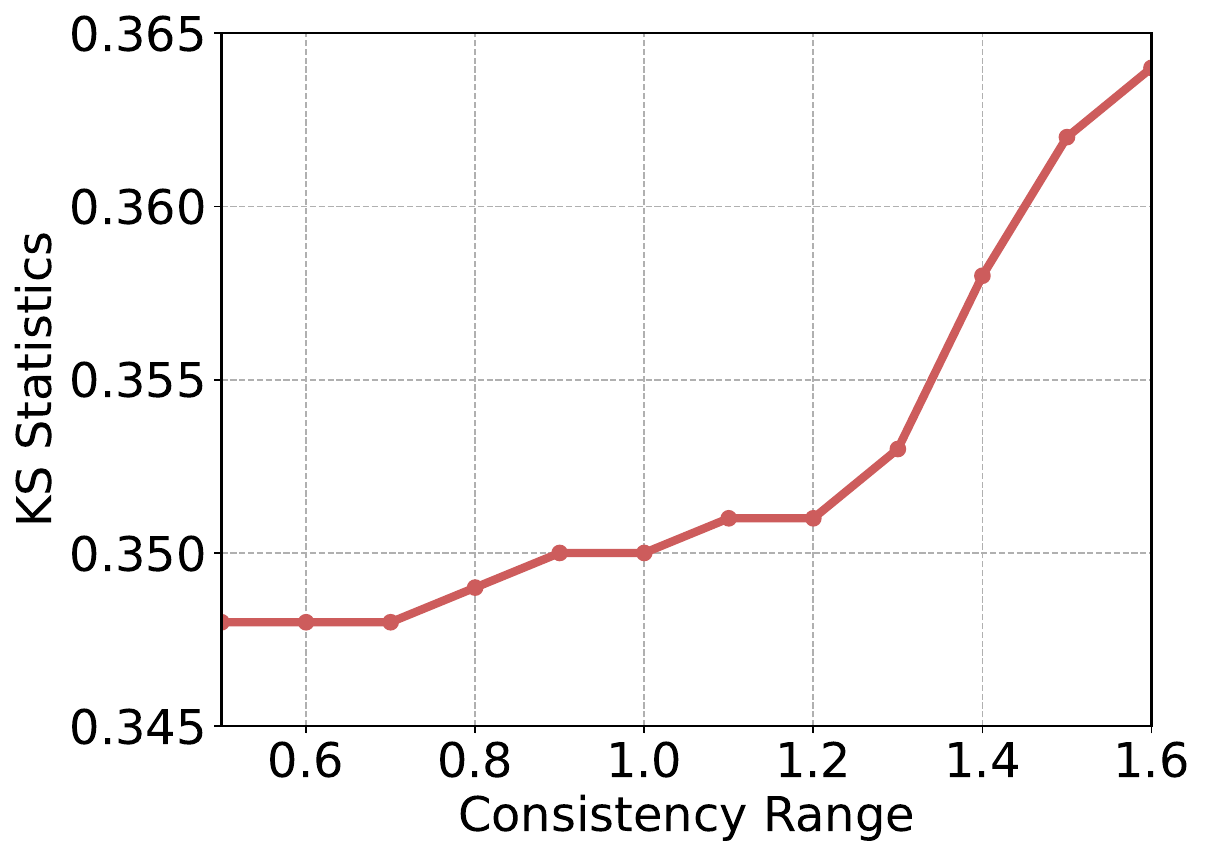}
  \vspace{-1mm}
  \caption{Trade-off between index correlation (left) and simulation quality (right).}
  \label{fig:cons_k}
  \Description{Trade-off between index correlation (left) and simulation quality (right).}
\end{figure}

As mentioned in Equation~(\ref{eq:adapt}), the hyper-parameter 
$k$ indicates the extent of influence that these indices may have on the output agent parameters, which eventually results in varying simulation qualities.
Figure \ref{fig:cons_k} illustrates this trade-off, suggesting that as the consistency range increases, the PCC steadily rises, while the simulation quality initially decreases slightly (until $k\approx 1.2$) and then drops drastically.
In the remaining experiments, we set $k=0.8$ to achieve a satisfactory correlation without significantly sacrificing simulation quality.

%1. consistency range - calibration quality + correlation - a trade-off: Figure~\ref{fig:cons_k}.

\begin{figure}[ht]
  \centering
  \includegraphics[width=70mm]{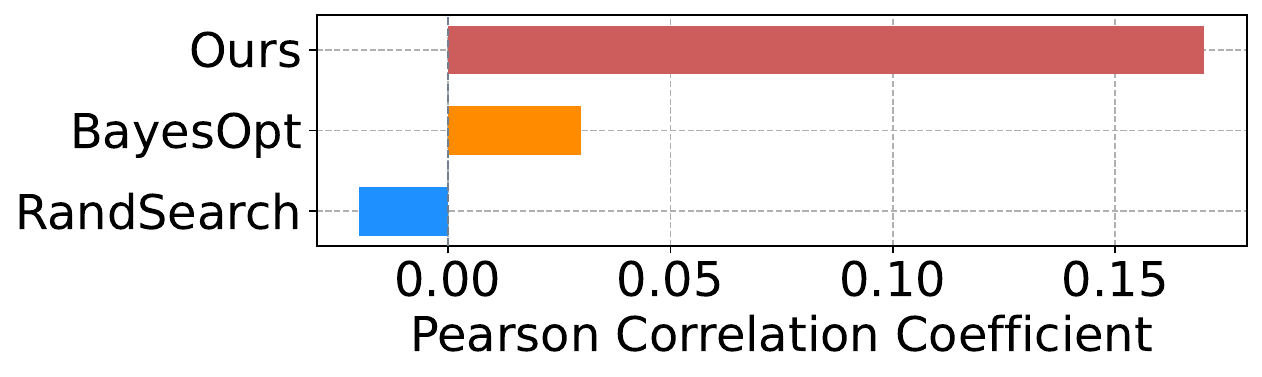}
  \vspace{-2mm}
  \caption{The chartist component under Market Trend.}
  \label{fig:schemes}
  \Description{The chartist component under Market Trend.}
\end{figure}

We also compare our methods with these two baselines in terms of the correlation coefficient.
Table~\ref{tab:corr_comparison} displays the average absolute PCCs of these indices with the calibrated agent parameters.
Our method demonstrates significantly higher correlation coefficients, as RandSearch and BayesOpt lack modules to deliberately correlate the agent parameters with these exogenous indices.
In particular, we visualize the specific PCCs of the \textit{chartist} component computed by each method based on the \textit{Market Trend} indices, which indicate the tendency of the market to move in a particular direction over a certain period,
as shown in Figure~\ref{fig:schemes}.
Understandably, when the market exhibits a relatively clear trend, the final trading decision should rely more on the decisions made by the \textit{chartist} component.

%2. overall correlation with indices, comparison: Table~\ref{tab:corr_comparison}.
%\textcolor{red}{used to be called behavioral variations, is it the same thing?}

\begin{table}[t]
\centering
\caption{Correlation between agent parameters and economic indices under different calibration methods.}
\label{tab:corr_comparison}
\begin{tabular}{@{}c|rrr@{}}
\toprule
\multicolumn{1}{l|}{} & RandSearch & BayesOpt & \textbf{Ours}   \\ \midrule
CPI                   & 0.0764     & 0.0447   & 0.142 \\ % 0.094 \\ % 0.2555 \\
PPI                   & 0.0790     & 0.0513   & 0.132 \\ % 0.032 \\ % 0.2595 \\
PMI                   & 0.0494     & 0.0549   & 0.146 \\ % -0.022 \\ % 0.2634 \\
Market Trend          & 0.0466     & 0.0921   & 0.158 \\ % -0.034 \\ % 0.3266 \\
Market Noise          & 0.0122     & 0.0601   & 0.134 \\ % 0.082 \\ % 0.3102 \\
\bottomrule
\end{tabular}
\end{table}

Finally, we present all the calibrated agent parameters under each category of indices in Table~\ref{tab:computed_corr}.
We aim to highlight some surprisingly reasonable results, as well as some failure cases.
\begin{enumerate}
	\item [\textcolor{teal}{\ding{51}}] High CPI typically indicates high inflation. In such circumstances, people tend to choose high-value assets and become more risk-averse to hedge against inflation. The first line of Table~\ref{tab:computed_corr} demonstrates a strong positive correlation with the fundamentalist ($\alpha_F$) and the risk aversion factor ($\beta_r$). People may also cautiously examine whether an asset is stable over a long period, hence a fair portion of the chartist component ($\alpha_C$) but with a relatively strong correlation with the horizon length ($\tau$).
	
	\item [\textcolor{purple}{\ding{55}}] PPI is usually correlated with CPI. The former is a metric from the producers' side, while the latter is a metric from the consumers' side. Therefore, they are expected to reflect at least some similar key features of the market. However, our system exhibits a negative correlation with the fundamentalist and the risk aversion factor.

	\item [\textcolor{teal}{\ding{51}}] High PMI normally presents an optimistic outlook on the current market, therefore, people may be more willing to take on additional risk in search of higher returns. This explains the negative correlation with the fundamentalist ($\alpha_F$) and the risk aversion factor ($\beta_r$), indicating that investors are, to some extent, risk-seeking. Also, in such circumstances, people may make homogeneous decisions regarding some emerging markets, resulting in a relatively stronger positive correlation with the chartist component ($\alpha_C$ and $\tau$).

	\item [\textcolor{teal}{\ding{51}}] The alignment with the index of Market Trend also leads to reasonable outcomes. Most traders tend to be trend-followers (showing a high  correlation with $\alpha_C$ and $\tau$), while possibly ignoring the true fundamental values of the assets (indicated by a negative correlation with $\alpha_F$), if a clear market trend is presented to the public.
	
	\item [\textcolor{purple}{\ding{55}}] Under the index of Market Noise, the positive correlation with $\sigma_N$ and $\beta_r$ as well as the negative correlation with $\alpha_C$ seem acceptable, whereas the positive correlation with $\alpha_F$ is not desirable as most traders may simply be irrational ones.
\end{enumerate}

%3. computed parameters with correlation: Table~\ref{tab:computed_corr}.

 \begin{table}[t]
    \centering
    \caption{Correlation coefficients between each economic index and agent parameters.}
    \label{tab:computed_corr}
\begin{tabular}{@{}c|rrrrr@{}}
\toprule
             & $\alpha_F$ & $\alpha_C$ & $\tau$ & $\sigma_N$ & $\beta_r$ \\ \midrule
CPI          & 0.17       & 0.13       & 0.14   & -0.12      & 0.15      \\
PPI          & -0.12      & 0.14       & 0.16   & 0.11       & -0.13     \\
PMI          & -0.12      & 0.16       & 0.15   & -0.15      & -0.15     \\
Market Trend & -0.15      & 0.17       & 0.14   & -0.14      & -0.19     \\
Market Noise & 0.13       & -0.13      & 0.12   & 0.12       & 0.17      \\ \bottomrule
\end{tabular}
\end{table}

%4. a closer look: only chartist under Market Trend: Figure~\ref{fig:schemes}. 
%\textcolor{red}{is it possible to plot the entire line of the second last line of Table~\ref{tab:computed_corr}.}

\section{Conclusion}

\label{sec:conclusion}

We present a formal game-theoretic framework along with an agent-based model (ABM) for studying the real-world stock market.
This ABM incorporates principles of behavioral economics to guide the decision-making of each trading agent.
In particular, we link order sizes to the quoted prices to reflect risk aversion.
The parameters of the agent model are first calibrated using a DNN-based approach and then correlated with publicly accessible economic indices.
This latter step enhances the explainability of our system. 
Through comprehensive experiments, we demonstrate the effectiveness (in terms of metrics such as price deviation and the Kolmogorov-Smirnov (KS) statistic) and efficiency (regarding both training and inference costs) of our approach compared to existing methods.
Several case studies are conducted to demonstrate that the agent parameters can be reasonably correlated with these economic indices.
We leave the specific usage of our simulator for devising investment strategies to future work.

%Limitation
%
%Future work
%
%1. Explicitly use this off-the-shelf simulator to devise investment strategies.

%1. neural order book (not truthful) like neural auction~\cite{dutting2024optimal}

%2. Re-formulation as One-Step MDPs, policy gradient, Boltzmann distribution 

%%%%%%%%%%%%%%%%%%%%%%%%%%%%%%%%%%%%%%%%%%%%%%%%%%%%%%%%%%%%%%%%%%%%%%%%

%%% The acknowledgments section is defined using the "acks" environment
%%% (rather than an unnumbered section). The use of this environment 
%%% ensures the proper identification of the section in the article 
%%% metadata as well as the consistent spelling of the heading.

%\begin{acks}
%If you wish to include any acknowledgments in your paper (e.g., to 
%people or funding agencies), please do so using the `\texttt{acks}' 
%environment. Note that the text of your acknowledgments will be omitted
%if you compile your document with the `\texttt{anonymous}' option.
%\end{acks}

%%%%%%%%%%%%%%%%%%%%%%%%%%%%%%%%%%%%%%%%%%%%%%%%%%%%%%%%%%%%%%%%%%%%%%%%

%%% The next two lines define, first, the bibliography style to be 
%%% applied, and, second, the bibliography file to be used.

%\clearpage

\bibliographystyle{ACM-Reference-Format} 
\bibliography{sample}


\begin{thebibliography}{49}

%%% the Web address.

\ifx \showCODEN    \undefined \def \showCODEN     #1{\unskip}     \fi
\ifx \showDOI      \undefined \def \showDOI       #1{#1}\fi
\ifx \showISBNx    \undefined \def \showISBNx     #1{\unskip}     \fi
\ifx \showISBNxiii \undefined \def \showISBNxiii  #1{\unskip}     \fi
\ifx \showISSN     \undefined \def \showISSN      #1{\unskip}     \fi
\ifx \showLCCN     \undefined \def \showLCCN      #1{\unskip}     \fi
\ifx \shownote     \undefined \def \shownote      #1{#1}          \fi
\ifx \showarticletitle \undefined \def \showarticletitle #1{#1}   \fi
\ifx \showURL      \undefined \def \showURL       {\relax}        \fi
% invisible to TeX
\providecommand\bibfield[2]{#2}
\providecommand\bibinfo[2]{#2}
\providecommand\natexlab[1]{#1}
\providecommand\showeprint[2][]{arXiv:#2}

\bibitem[\protect\citeauthoryear{Bai, Cui, Zhang, Xu, Wang, and Hancock}{Bai et~al\mbox{.}}{2020}]%
        {bai2020entropic}
\bibfield{author}{\bibinfo{person}{Lu Bai}, \bibinfo{person}{Lixin Cui}, \bibinfo{person}{Zhihong Zhang}, \bibinfo{person}{Lixiang Xu}, \bibinfo{person}{Yue Wang}, {and} \bibinfo{person}{Edwin~R Hancock}.} \bibinfo{year}{2020}\natexlab{}.
\newblock \showarticletitle{Entropic dynamic time warping kernels for co-evolving financial time series analysis}.
\newblock \bibinfo{journal}{\emph{IEEE Transactions on Neural Networks and Learning Systems}} \bibinfo{volume}{34}, \bibinfo{number}{4} (\bibinfo{year}{2020}), \bibinfo{pages}{1808--1822}.
\newblock


\bibitem[\protect\citeauthoryear{Bai, Lam, Balch, and Vyetrenko}{Bai et~al\mbox{.}}{2022}]%
        {bai2022efficient}
\bibfield{author}{\bibinfo{person}{Yuanlu Bai}, \bibinfo{person}{Henry Lam}, \bibinfo{person}{Tucker Balch}, {and} \bibinfo{person}{Svitlana Vyetrenko}.} \bibinfo{year}{2022}\natexlab{}.
\newblock \showarticletitle{Efficient calibration of multi-agent simulation models from output series with bayesian optimization}. In \bibinfo{booktitle}{\emph{Proceedings of the Third ACM International Conference on AI in Finance}}. \bibinfo{pages}{437--445}.
\newblock


\bibitem[\protect\citeauthoryear{Bak, Paczuski, and Shubik}{Bak et~al\mbox{.}}{1997}]%
        {bak1997price}
\bibfield{author}{\bibinfo{person}{Per Bak}, \bibinfo{person}{Maya Paczuski}, {and} \bibinfo{person}{Martin Shubik}.} \bibinfo{year}{1997}\natexlab{}.
\newblock \showarticletitle{Price variations in a stock market with many agents}.
\newblock \bibinfo{journal}{\emph{Physica A: Statistical Mechanics and its Applications}} \bibinfo{volume}{246}, \bibinfo{number}{3-4} (\bibinfo{year}{1997}), \bibinfo{pages}{430--453}.
\newblock


\bibitem[\protect\citeauthoryear{Bouchaud, Bonart, Donier, and Gould}{Bouchaud et~al\mbox{.}}{2018}]%
        {bouchaud2018trades}
\bibfield{author}{\bibinfo{person}{Jean-Philippe Bouchaud}, \bibinfo{person}{Julius Bonart}, \bibinfo{person}{Jonathan Donier}, {and} \bibinfo{person}{Martin Gould}.} \bibinfo{year}{2018}\natexlab{}.
\newblock \bibinfo{booktitle}{\emph{Trades, quotes and prices: financial markets under the microscope}}.
\newblock \bibinfo{publisher}{Cambridge University Press}.
\newblock


\bibitem[\protect\citeauthoryear{Bouchaud, M{\'e}zard, and Potters}{Bouchaud et~al\mbox{.}}{2002}]%
        {bouchaud2002statistical}
\bibfield{author}{\bibinfo{person}{Jean-Philippe Bouchaud}, \bibinfo{person}{Marc M{\'e}zard}, {and} \bibinfo{person}{Marc Potters}.} \bibinfo{year}{2002}\natexlab{}.
\newblock \showarticletitle{Statistical properties of stock order books: empirical results and models}.
\newblock \bibinfo{journal}{\emph{Quantitative finance}} \bibinfo{volume}{2}, \bibinfo{number}{4} (\bibinfo{year}{2002}), \bibinfo{pages}{251}.
\newblock


\bibitem[\protect\citeauthoryear{Brogaard and Zareei}{Brogaard and Zareei}{2023}]%
        {brogaard2023machine}
\bibfield{author}{\bibinfo{person}{Jonathan Brogaard} {and} \bibinfo{person}{Abalfazl Zareei}.} \bibinfo{year}{2023}\natexlab{}.
\newblock \showarticletitle{Machine learning and the stock market}.
\newblock \bibinfo{journal}{\emph{Journal of Financial and Quantitative Analysis}} \bibinfo{volume}{58}, \bibinfo{number}{4} (\bibinfo{year}{2023}), \bibinfo{pages}{1431--1472}.
\newblock


\bibitem[\protect\citeauthoryear{Byrd, Hybinette, and Balch}{Byrd et~al\mbox{.}}{2019}]%
        {byrd2019abides}
\bibfield{author}{\bibinfo{person}{David Byrd}, \bibinfo{person}{Maria Hybinette}, {and} \bibinfo{person}{Tucker~Hybinette Balch}.} \bibinfo{year}{2019}\natexlab{}.
\newblock \showarticletitle{Abides: Towards high-fidelity market simulation for ai research}.
\newblock \bibinfo{journal}{\emph{arXiv preprint arXiv:1904.12066}} (\bibinfo{year}{2019}).
\newblock


\bibitem[\protect\citeauthoryear{Chiarella}{Chiarella}{1992}]%
        {chiarella1992dynamics}
\bibfield{author}{\bibinfo{person}{Carl Chiarella}.} \bibinfo{year}{1992}\natexlab{}.
\newblock \showarticletitle{The dynamics of speculative behaviour}.
\newblock \bibinfo{journal}{\emph{Annals of operations research}} \bibinfo{volume}{37}, \bibinfo{number}{1} (\bibinfo{year}{1992}), \bibinfo{pages}{101--123}.
\newblock


\bibitem[\protect\citeauthoryear{Chiarella, Dieci, and Gardini}{Chiarella et~al\mbox{.}}{2006}]%
        {chiarella2006asset}
\bibfield{author}{\bibinfo{person}{Carl Chiarella}, \bibinfo{person}{Roberto Dieci}, {and} \bibinfo{person}{Laura Gardini}.} \bibinfo{year}{2006}\natexlab{}.
\newblock \showarticletitle{Asset price and wealth dynamics in a financial market with heterogeneous agents}.
\newblock \bibinfo{journal}{\emph{Journal of Economic Dynamics and Control}} \bibinfo{volume}{30}, \bibinfo{number}{9-10} (\bibinfo{year}{2006}), \bibinfo{pages}{1755--1786}.
\newblock


\bibitem[\protect\citeauthoryear{Chiarella, Iori, and Perell{\'o}}{Chiarella et~al\mbox{.}}{2009}]%
        {chiarella2009impact}
\bibfield{author}{\bibinfo{person}{Carl Chiarella}, \bibinfo{person}{Giulia Iori}, {and} \bibinfo{person}{Josep Perell{\'o}}.} \bibinfo{year}{2009}\natexlab{}.
\newblock \showarticletitle{The impact of heterogeneous trading rules on the limit order book and order flows}.
\newblock \bibinfo{journal}{\emph{Journal of Economic Dynamics and Control}} \bibinfo{volume}{33}, \bibinfo{number}{3} (\bibinfo{year}{2009}), \bibinfo{pages}{525--537}.
\newblock


\bibitem[\protect\citeauthoryear{Coletta, Moulin, Vyetrenko, and Balch}{Coletta et~al\mbox{.}}{2022}]%
        {coletta2022learning}
\bibfield{author}{\bibinfo{person}{Andrea Coletta}, \bibinfo{person}{Aymeric Moulin}, \bibinfo{person}{Svitlana Vyetrenko}, {and} \bibinfo{person}{Tucker Balch}.} \bibinfo{year}{2022}\natexlab{}.
\newblock \showarticletitle{Learning to simulate realistic limit order book markets from data as a World Agent}. In \bibinfo{booktitle}{\emph{Proceedings of the Third ACM International Conference on AI in Finance}}. \bibinfo{pages}{428--436}.
\newblock


\bibitem[\protect\citeauthoryear{Coletta, Prata, Conti, Mercanti, Bartolini, Moulin, Vyetrenko, and Balch}{Coletta et~al\mbox{.}}{2021}]%
        {coletta2021towards}
\bibfield{author}{\bibinfo{person}{Andrea Coletta}, \bibinfo{person}{Matteo Prata}, \bibinfo{person}{Michele Conti}, \bibinfo{person}{Emanuele Mercanti}, \bibinfo{person}{Novella Bartolini}, \bibinfo{person}{Aymeric Moulin}, \bibinfo{person}{Svitlana Vyetrenko}, {and} \bibinfo{person}{Tucker Balch}.} \bibinfo{year}{2021}\natexlab{}.
\newblock \showarticletitle{Towards realistic market simulations: a generative adversarial networks approach}. In \bibinfo{booktitle}{\emph{Proceedings of the Second ACM International Conference on AI in Finance}}. \bibinfo{pages}{1--9}.
\newblock


\bibitem[\protect\citeauthoryear{Deng, Bao, Kong, Ren, and Dai}{Deng et~al\mbox{.}}{2016}]%
        {deng2016deep}
\bibfield{author}{\bibinfo{person}{Yue Deng}, \bibinfo{person}{Feng Bao}, \bibinfo{person}{Youyong Kong}, \bibinfo{person}{Zhiquan Ren}, {and} \bibinfo{person}{Qionghai Dai}.} \bibinfo{year}{2016}\natexlab{}.
\newblock \showarticletitle{Deep direct reinforcement learning for financial signal representation and trading}.
\newblock \bibinfo{journal}{\emph{IEEE transactions on neural networks and learning systems}} \bibinfo{volume}{28}, \bibinfo{number}{3} (\bibinfo{year}{2016}), \bibinfo{pages}{653--664}.
\newblock


\bibitem[\protect\citeauthoryear{Friedman}{Friedman}{2018}]%
        {friedman2018double}
\bibfield{author}{\bibinfo{person}{Daniel Friedman}.} \bibinfo{year}{2018}\natexlab{}.
\newblock \showarticletitle{The double auction market institution: A survey}.
\newblock In \bibinfo{booktitle}{\emph{The double auction market}}. \bibinfo{publisher}{Routledge}, \bibinfo{pages}{3--26}.
\newblock


\bibitem[\protect\citeauthoryear{Gjerstad}{Gjerstad}{2007}]%
        {gjerstad2007competitive}
\bibfield{author}{\bibinfo{person}{Steven Gjerstad}.} \bibinfo{year}{2007}\natexlab{}.
\newblock \showarticletitle{The competitive market paradox}.
\newblock \bibinfo{journal}{\emph{Journal of Economic Dynamics and Control}} \bibinfo{volume}{31}, \bibinfo{number}{5} (\bibinfo{year}{2007}), \bibinfo{pages}{1753--1780}.
\newblock


\bibitem[\protect\citeauthoryear{Gjerstad and Dickhaut}{Gjerstad and Dickhaut}{1998}]%
        {gjerstad1998price}
\bibfield{author}{\bibinfo{person}{Steven Gjerstad} {and} \bibinfo{person}{John Dickhaut}.} \bibinfo{year}{1998}\natexlab{}.
\newblock \showarticletitle{Price formation in double auctions}.
\newblock \bibinfo{journal}{\emph{Games and economic behavior}} \bibinfo{volume}{22}, \bibinfo{number}{1} (\bibinfo{year}{1998}), \bibinfo{pages}{1--29}.
\newblock


\bibitem[\protect\citeauthoryear{Gould, Porter, Williams, McDonald, Fenn, and Howison}{Gould et~al\mbox{.}}{2013}]%
        {gould2013limit}
\bibfield{author}{\bibinfo{person}{Martin~D Gould}, \bibinfo{person}{Mason~A Porter}, \bibinfo{person}{Stacy Williams}, \bibinfo{person}{Mark McDonald}, \bibinfo{person}{Daniel~J Fenn}, {and} \bibinfo{person}{Sam~D Howison}.} \bibinfo{year}{2013}\natexlab{}.
\newblock \showarticletitle{Limit order books}.
\newblock \bibinfo{journal}{\emph{Quantitative Finance}} \bibinfo{volume}{13}, \bibinfo{number}{11} (\bibinfo{year}{2013}), \bibinfo{pages}{1709--1742}.
\newblock


\bibitem[\protect\citeauthoryear{G{\"u}lmez}{G{\"u}lmez}{2023}]%
        {gulmez2023stock}
\bibfield{author}{\bibinfo{person}{Burak G{\"u}lmez}.} \bibinfo{year}{2023}\natexlab{}.
\newblock \showarticletitle{Stock price prediction with optimized deep LSTM network with artificial rabbits optimization algorithm}.
\newblock \bibinfo{journal}{\emph{Expert Systems with Applications}}  \bibinfo{volume}{227} (\bibinfo{year}{2023}), \bibinfo{pages}{120346}.
\newblock


\bibitem[\protect\citeauthoryear{Han, Kim, and Enke}{Han et~al\mbox{.}}{2023}]%
        {han2023machine}
\bibfield{author}{\bibinfo{person}{Yechan Han}, \bibinfo{person}{Jaeyun Kim}, {and} \bibinfo{person}{David Enke}.} \bibinfo{year}{2023}\natexlab{}.
\newblock \showarticletitle{A machine learning trading system for the stock market based on N-period Min-Max labeling using XGBoost}.
\newblock \bibinfo{journal}{\emph{Expert Systems with Applications}}  \bibinfo{volume}{211} (\bibinfo{year}{2023}), \bibinfo{pages}{118581}.
\newblock


\bibitem[\protect\citeauthoryear{Hansen, Bernstein, and Zilberstein}{Hansen et~al\mbox{.}}{2004}]%
        {hansen2004dynamic}
\bibfield{author}{\bibinfo{person}{Eric~A Hansen}, \bibinfo{person}{Daniel~S Bernstein}, {and} \bibinfo{person}{Shlomo Zilberstein}.} \bibinfo{year}{2004}\natexlab{}.
\newblock \showarticletitle{Dynamic programming for partially observable stochastic games}. In \bibinfo{booktitle}{\emph{AAAI}}, Vol.~\bibinfo{volume}{4}. \bibinfo{pages}{709--715}.
\newblock


\bibitem[\protect\citeauthoryear{Hochreiter and Schmidhuber}{Hochreiter and Schmidhuber}{1997}]%
        {hochreiter1997long}
\bibfield{author}{\bibinfo{person}{Sepp Hochreiter} {and} \bibinfo{person}{J{\"u}rgen Schmidhuber}.} \bibinfo{year}{1997}\natexlab{}.
\newblock \showarticletitle{Long short-term memory}.
\newblock \bibinfo{journal}{\emph{Neural computation}} \bibinfo{volume}{9}, \bibinfo{number}{8} (\bibinfo{year}{1997}), \bibinfo{pages}{1735--1780}.
\newblock


\bibitem[\protect\citeauthoryear{Hou, Xu, Li, Liu, Liu, Chen, and Bian}{Hou et~al\mbox{.}}{2022}]%
        {hou2022multi}
\bibfield{author}{\bibinfo{person}{Min Hou}, \bibinfo{person}{Chang Xu}, \bibinfo{person}{Zhi Li}, \bibinfo{person}{Yang Liu}, \bibinfo{person}{Weiqing Liu}, \bibinfo{person}{Enhong Chen}, {and} \bibinfo{person}{Jiang Bian}.} \bibinfo{year}{2022}\natexlab{}.
\newblock \showarticletitle{Multi-Granularity Residual Learning with Confidence Estimation for Time Series Prediction}. In \bibinfo{booktitle}{\emph{Proceedings of the ACM Web Conference 2022}}. \bibinfo{pages}{112--121}.
\newblock


\bibitem[\protect\citeauthoryear{Hou, Xu, Liu, Liu, Bian, Wu, Li, Chen, and Liu}{Hou et~al\mbox{.}}{2021}]%
        {hou2021stock}
\bibfield{author}{\bibinfo{person}{Min Hou}, \bibinfo{person}{Chang Xu}, \bibinfo{person}{Yang Liu}, \bibinfo{person}{Weiqing Liu}, \bibinfo{person}{Jiang Bian}, \bibinfo{person}{Le Wu}, \bibinfo{person}{Zhi Li}, \bibinfo{person}{Enhong Chen}, {and} \bibinfo{person}{Tie-Yan Liu}.} \bibinfo{year}{2021}\natexlab{}.
\newblock \showarticletitle{Stock trend prediction with multi-granularity data: A contrastive learning approach with adaptive fusion}. In \bibinfo{booktitle}{\emph{Proceedings of the 30th ACM International Conference on Information \& Knowledge Management}}. \bibinfo{pages}{700--709}.
\newblock


\bibitem[\protect\citeauthoryear{Hsu, Tsai, and Li}{Hsu et~al\mbox{.}}{2021}]%
        {hsu2021fingat}
\bibfield{author}{\bibinfo{person}{Yi-Ling Hsu}, \bibinfo{person}{Yu-Che Tsai}, {and} \bibinfo{person}{Cheng-Te Li}.} \bibinfo{year}{2021}\natexlab{}.
\newblock \showarticletitle{FinGAT: Financial graph attention networks for recommending top-$ k $ k profitable stocks}.
\newblock \bibinfo{journal}{\emph{IEEE transactions on knowledge and data engineering}} \bibinfo{volume}{35}, \bibinfo{number}{1} (\bibinfo{year}{2021}), \bibinfo{pages}{469--481}.
\newblock


\bibitem[\protect\citeauthoryear{Karpe, Fang, Ma, and Wang}{Karpe et~al\mbox{.}}{2020}]%
        {karpe2020multi}
\bibfield{author}{\bibinfo{person}{Micha{\"e}l Karpe}, \bibinfo{person}{Jin Fang}, \bibinfo{person}{Zhongyao Ma}, {and} \bibinfo{person}{Chen Wang}.} \bibinfo{year}{2020}\natexlab{}.
\newblock \showarticletitle{Multi-agent reinforcement learning in a realistic limit order book market simulation}. In \bibinfo{booktitle}{\emph{Proceedings of the First ACM International Conference on AI in Finance}}. \bibinfo{pages}{1--7}.
\newblock


\bibitem[\protect\citeauthoryear{Kaufman}{Kaufman}{2013}]%
        {kaufman2013trading}
\bibfield{author}{\bibinfo{person}{Perry~J Kaufman}.} \bibinfo{year}{2013}\natexlab{}.
\newblock \bibinfo{booktitle}{\emph{Trading Systems and Methods}}.
\newblock \bibinfo{publisher}{John Wiley \& Sons}.
\newblock


\bibitem[\protect\citeauthoryear{Kingma and Welling}{Kingma and Welling}{2013}]%
        {kingma2013auto}
\bibfield{author}{\bibinfo{person}{Diederik~P Kingma} {and} \bibinfo{person}{Max Welling}.} \bibinfo{year}{2013}\natexlab{}.
\newblock \showarticletitle{Auto-encoding variational bayes}.
\newblock \bibinfo{journal}{\emph{arXiv preprint arXiv:1312.6114}} (\bibinfo{year}{2013}).
\newblock


\bibitem[\protect\citeauthoryear{LeBaron, Arthur, and Palmer}{LeBaron et~al\mbox{.}}{1999}]%
        {lebaron1999time}
\bibfield{author}{\bibinfo{person}{Blake LeBaron}, \bibinfo{person}{W~Brian Arthur}, {and} \bibinfo{person}{Richard Palmer}.} \bibinfo{year}{1999}\natexlab{}.
\newblock \showarticletitle{Time series properties of an artificial stock market}.
\newblock \bibinfo{journal}{\emph{Journal of Economic Dynamics and control}} \bibinfo{volume}{23}, \bibinfo{number}{9-10} (\bibinfo{year}{1999}), \bibinfo{pages}{1487--1516}.
\newblock


\bibitem[\protect\citeauthoryear{Li, Wang, Lin, Sinha, and Wellman}{Li et~al\mbox{.}}{2020}]%
        {li2020generating}
\bibfield{author}{\bibinfo{person}{Junyi Li}, \bibinfo{person}{Xintong Wang}, \bibinfo{person}{Yaoyang Lin}, \bibinfo{person}{Arunesh Sinha}, {and} \bibinfo{person}{Michael Wellman}.} \bibinfo{year}{2020}\natexlab{}.
\newblock \showarticletitle{Generating realistic stock market order streams}. In \bibinfo{booktitle}{\emph{Proceedings of the AAAI Conference on Artificial Intelligence}}, Vol.~\bibinfo{volume}{34}. \bibinfo{pages}{727--734}.
\newblock


\bibitem[\protect\citeauthoryear{Li and Das}{Li and Das}{2016}]%
        {li2016agent}
\bibfield{author}{\bibinfo{person}{Zhuoshu Li} {and} \bibinfo{person}{Sanmay Das}.} \bibinfo{year}{2016}\natexlab{}.
\newblock \showarticletitle{An agent-based model of competition between financial exchanges: Can frequent call mechanisms drive trade away from CDAs?}. In \bibinfo{booktitle}{\emph{Proceedings of the 2016 International Conference on Autonomous Agents \& Multiagent Systems}}. \bibinfo{pages}{50--58}.
\newblock


\bibitem[\protect\citeauthoryear{Liu, Rui, Gao, Yang, Yang, Wang, Wang, and Guo}{Liu et~al\mbox{.}}{2021}]%
        {liu2021finrl}
\bibfield{author}{\bibinfo{person}{Xiao-Yang Liu}, \bibinfo{person}{Jingyang Rui}, \bibinfo{person}{Jiechao Gao}, \bibinfo{person}{Liuqing Yang}, \bibinfo{person}{Hongyang Yang}, \bibinfo{person}{Zhaoran Wang}, \bibinfo{person}{Christina~Dan Wang}, {and} \bibinfo{person}{Jian Guo}.} \bibinfo{year}{2021}\natexlab{}.
\newblock \showarticletitle{FinRL-meta: A universe of near-real market environments for data-driven deep reinforcement learning in quantitative finance}.
\newblock \bibinfo{journal}{\emph{arXiv preprint arXiv:2112.06753}} (\bibinfo{year}{2021}).
\newblock


\bibitem[\protect\citeauthoryear{Majewski, Ciliberti, and Bouchaud}{Majewski et~al\mbox{.}}{2020}]%
        {majewski2020co}
\bibfield{author}{\bibinfo{person}{Adam~A Majewski}, \bibinfo{person}{Stefano Ciliberti}, {and} \bibinfo{person}{Jean-Philippe Bouchaud}.} \bibinfo{year}{2020}\natexlab{}.
\newblock \showarticletitle{Co-existence of trend and value in financial markets: Estimating an extended Chiarella model}.
\newblock \bibinfo{journal}{\emph{Journal of Economic Dynamics and Control}}  \bibinfo{volume}{112} (\bibinfo{year}{2020}), \bibinfo{pages}{103791}.
\newblock


\bibitem[\protect\citeauthoryear{Maslov}{Maslov}{2000}]%
        {maslov2000simple}
\bibfield{author}{\bibinfo{person}{Sergei Maslov}.} \bibinfo{year}{2000}\natexlab{}.
\newblock \showarticletitle{Simple model of a limit order-driven market}.
\newblock \bibinfo{journal}{\emph{Physica A: Statistical Mechanics and its Applications}} \bibinfo{volume}{278}, \bibinfo{number}{3-4} (\bibinfo{year}{2000}), \bibinfo{pages}{571--578}.
\newblock


\bibitem[\protect\citeauthoryear{Mi, Xia, Song, Zhang, Zhu, and Wang}{Mi et~al\mbox{.}}{2024}]%
        {mi2024taxai}
\bibfield{author}{\bibinfo{person}{Qirui Mi}, \bibinfo{person}{Siyu Xia}, \bibinfo{person}{Yan Song}, \bibinfo{person}{Haifeng Zhang}, \bibinfo{person}{Shenghao Zhu}, {and} \bibinfo{person}{Jun Wang}.} \bibinfo{year}{2024}\natexlab{}.
\newblock \showarticletitle{TaxAI: A Dynamic Economic Simulator and Benchmark for Multi-agent Reinforcement Learning}. In \bibinfo{booktitle}{\emph{Proceedings of the 23rd International Conference on Autonomous Agents and Multiagent Systems}}. \bibinfo{pages}{1390--1399}.
\newblock


\bibitem[\protect\citeauthoryear{Mi, Yang, Fan, Fan, Ma, Ma, Xia, An, Wang, and Zhang}{Mi et~al\mbox{.}}{2025}]%
        {mi2025econgym}
\bibfield{author}{\bibinfo{person}{Qirui Mi}, \bibinfo{person}{Qipeng Yang}, \bibinfo{person}{Zijun Fan}, \bibinfo{person}{Wentian Fan}, \bibinfo{person}{Heyang Ma}, \bibinfo{person}{Chengdong Ma}, \bibinfo{person}{Siyu Xia}, \bibinfo{person}{Bo An}, \bibinfo{person}{Jun Wang}, {and} \bibinfo{person}{Haifeng Zhang}.} \bibinfo{year}{2025}\natexlab{}.
\newblock \showarticletitle{EconGym: A Scalable AI Testbed with Diverse Economic Tasks}.
\newblock \bibinfo{journal}{\emph{arXiv preprint arXiv:2506.12110}} (\bibinfo{year}{2025}).
\newblock


\bibitem[\protect\citeauthoryear{Paddrik, Hayes, Todd, Yang, Beling, and Scherer}{Paddrik et~al\mbox{.}}{2012}]%
        {paddrik2012agent}
\bibfield{author}{\bibinfo{person}{Mark Paddrik}, \bibinfo{person}{Roy Hayes}, \bibinfo{person}{Andrew Todd}, \bibinfo{person}{Steve Yang}, \bibinfo{person}{Peter Beling}, {and} \bibinfo{person}{William Scherer}.} \bibinfo{year}{2012}\natexlab{}.
\newblock \showarticletitle{An agent based model of the E-Mini S\&P 500 applied to Flash Crash analysis}. In \bibinfo{booktitle}{\emph{2012 IEEE conference on computational intelligence for financial engineering \& economics (CIFEr)}}. IEEE, \bibinfo{pages}{1--8}.
\newblock


\bibitem[\protect\citeauthoryear{Shapley}{Shapley}{1953}]%
        {shapley1953stochastic}
\bibfield{author}{\bibinfo{person}{Lloyd~S Shapley}.} \bibinfo{year}{1953}\natexlab{}.
\newblock \showarticletitle{Stochastic games}.
\newblock \bibinfo{journal}{\emph{Proceedings of the national academy of sciences}} \bibinfo{volume}{39}, \bibinfo{number}{10} (\bibinfo{year}{1953}), \bibinfo{pages}{1095--1100}.
\newblock


\bibitem[\protect\citeauthoryear{Shi and Cartlidge}{Shi and Cartlidge}{2022}]%
        {shi2022state}
\bibfield{author}{\bibinfo{person}{Zijian Shi} {and} \bibinfo{person}{John Cartlidge}.} \bibinfo{year}{2022}\natexlab{}.
\newblock \showarticletitle{State Dependent Parallel Neural Hawkes Process for Limit Order Book Event Stream Prediction and Simulation}. In \bibinfo{booktitle}{\emph{Proceedings of the 28th ACM SIGKDD Conference on Knowledge Discovery and Data Mining}}. \bibinfo{pages}{1607--1615}.
\newblock


\bibitem[\protect\citeauthoryear{Stillman, Baggott, Lyon, Zhang, Zhu, Chen, and Vytelingum}{Stillman et~al\mbox{.}}{2023}]%
        {stillman2023deep}
\bibfield{author}{\bibinfo{person}{Namid~R Stillman}, \bibinfo{person}{Rory Baggott}, \bibinfo{person}{Justin Lyon}, \bibinfo{person}{Jianfei Zhang}, \bibinfo{person}{Dingqui Zhu}, \bibinfo{person}{Tao Chen}, {and} \bibinfo{person}{Perukrishnen Vytelingum}.} \bibinfo{year}{2023}\natexlab{}.
\newblock \showarticletitle{Deep Calibration of Market Simulations using Neural Density Estimators and Embedding Networks}. In \bibinfo{booktitle}{\emph{Proceedings of the Fourth ACM International Conference on AI in Finance}}. \bibinfo{pages}{46--54}.
\newblock


\bibitem[\protect\citeauthoryear{Storchan, Vyetrenko, and Balch}{Storchan et~al\mbox{.}}{2021}]%
        {storchan2021learning}
\bibfield{author}{\bibinfo{person}{Victor Storchan}, \bibinfo{person}{Svitlana Vyetrenko}, {and} \bibinfo{person}{Tucker Balch}.} \bibinfo{year}{2021}\natexlab{}.
\newblock \showarticletitle{Learning who is in the market from time series: market participant discovery through adversarial calibration of multi-agent simulators}.
\newblock \bibinfo{journal}{\emph{arXiv preprint arXiv:2108.00664}} (\bibinfo{year}{2021}).
\newblock


\bibitem[\protect\citeauthoryear{Su, Huo, Zhang, Dou, Yu, Xu, Lu, and Zheng}{Su et~al\mbox{.}}{2024}]%
        {su2024auctionnet}
\bibfield{author}{\bibinfo{person}{Kefan Su}, \bibinfo{person}{Yusen Huo}, \bibinfo{person}{Zhilin Zhang}, \bibinfo{person}{Shuai Dou}, \bibinfo{person}{Chuan Yu}, \bibinfo{person}{Jian Xu}, \bibinfo{person}{Zongqing Lu}, {and} \bibinfo{person}{Bo Zheng}.} \bibinfo{year}{2024}\natexlab{}.
\newblock \showarticletitle{Auctionnet: A novel benchmark for decision-making in large-scale games}.
\newblock \bibinfo{journal}{\emph{Advances in Neural Information Processing Systems}}  \bibinfo{volume}{37} (\bibinfo{year}{2024}), \bibinfo{pages}{94428--94452}.
\newblock


\bibitem[\protect\citeauthoryear{Tesauro and Bredin}{Tesauro and Bredin}{2002}]%
        {tesauro2002strategic}
\bibfield{author}{\bibinfo{person}{Gerald Tesauro} {and} \bibinfo{person}{Jonathan~L Bredin}.} \bibinfo{year}{2002}\natexlab{}.
\newblock \showarticletitle{Strategic sequential bidding in auctions using dynamic programming}. In \bibinfo{booktitle}{\emph{Proceedings of the first international joint conference on Autonomous agents and multiagent systems: part 2}}. \bibinfo{pages}{591--598}.
\newblock


\bibitem[\protect\citeauthoryear{Tesauro and Das}{Tesauro and Das}{2001}]%
        {tesauro2001high}
\bibfield{author}{\bibinfo{person}{Gerald Tesauro} {and} \bibinfo{person}{Rajarshi Das}.} \bibinfo{year}{2001}\natexlab{}.
\newblock \showarticletitle{High-performance bidding agents for the continuous double auction}. In \bibinfo{booktitle}{\emph{Proceedings of the 3rd ACM Conference on Electronic Commerce}}. \bibinfo{pages}{206--209}.
\newblock


\bibitem[\protect\citeauthoryear{Vyetrenko, Byrd, Petosa, Mahfouz, Dervovic, Veloso, and Balch}{Vyetrenko et~al\mbox{.}}{2020}]%
        {vyetrenko2020get}
\bibfield{author}{\bibinfo{person}{Svitlana Vyetrenko}, \bibinfo{person}{David Byrd}, \bibinfo{person}{Nick Petosa}, \bibinfo{person}{Mahmoud Mahfouz}, \bibinfo{person}{Danial Dervovic}, \bibinfo{person}{Manuela Veloso}, {and} \bibinfo{person}{Tucker Balch}.} \bibinfo{year}{2020}\natexlab{}.
\newblock \showarticletitle{Get real: Realism metrics for robust limit order book market simulations}. In \bibinfo{booktitle}{\emph{Proceedings of the First ACM International Conference on AI in Finance}}. \bibinfo{pages}{1--8}.
\newblock


\bibitem[\protect\citeauthoryear{Wah and Wellman}{Wah and Wellman}{2017}]%
        {wah2017latency}
\bibfield{author}{\bibinfo{person}{Elaine Wah} {and} \bibinfo{person}{Michael~P Wellman}.} \bibinfo{year}{2017}\natexlab{}.
\newblock \showarticletitle{Latency arbitrage in fragmented markets: A strategic agent-based analysis}.
\newblock \bibinfo{journal}{\emph{Algorithmic Finance}} \bibinfo{volume}{5}, \bibinfo{number}{3-4} (\bibinfo{year}{2017}), \bibinfo{pages}{69--93}.
\newblock


\bibitem[\protect\citeauthoryear{Wang and Wellman}{Wang and Wellman}{2017}]%
        {wang2017spoofing}
\bibfield{author}{\bibinfo{person}{Xintong Wang} {and} \bibinfo{person}{Michael~P Wellman}.} \bibinfo{year}{2017}\natexlab{}.
\newblock \showarticletitle{Spoofing the Limit Order Book: An Agent-Based Model}. In \bibinfo{booktitle}{\emph{Proceedings of the 16th Conference on Autonomous Agents and MultiAgent Systems}}. \bibinfo{pages}{651--659}.
\newblock


\bibitem[\protect\citeauthoryear{Wilder}{Wilder}{1978}]%
        {wilder1978new}
\bibfield{author}{\bibinfo{person}{J~Welles Wilder}.} \bibinfo{year}{1978}\natexlab{}.
\newblock \bibinfo{booktitle}{\emph{New concepts in technical trading systems}}.
\newblock \bibinfo{publisher}{Greensboro, NC}.
\newblock


\bibitem[\protect\citeauthoryear{Xing, Cheng, Huang, Li, and Zhao}{Xing et~al\mbox{.}}{2023}]%
        {xing2023learning}
\bibfield{author}{\bibinfo{person}{Rong Xing}, \bibinfo{person}{Rui Cheng}, \bibinfo{person}{Jiwen Huang}, \bibinfo{person}{Qing Li}, {and} \bibinfo{person}{Jingmei Zhao}.} \bibinfo{year}{2023}\natexlab{}.
\newblock \showarticletitle{Learning to Understand the Vague Graph for Stock Prediction With Momentum Spillovers}.
\newblock \bibinfo{journal}{\emph{IEEE Transactions on Knowledge and Data Engineering}} (\bibinfo{year}{2023}).
\newblock


\bibitem[\protect\citeauthoryear{Zhu and Lin}{Zhu and Lin}{2025}]%
        {zhu2025single}
\bibfield{author}{\bibinfo{person}{Fengming Zhu} {and} \bibinfo{person}{Fangzhen Lin}.} \bibinfo{year}{2025}\natexlab{}.
\newblock \showarticletitle{Single-Agent Planning in a Multi-Agent System: A Unified Framework for Type-Based Planners}. In \bibinfo{booktitle}{\emph{Proceedings of the 24th International Conference on Autonomous Agents and Multiagent Systems}}. \bibinfo{pages}{2382--2391}.
\newblock


\end{thebibliography}

%%%%%%%%%%%%%%%%%%%%%%%%%%%%%%%%%%%%%%%%%%%%%%%%%%%%%%%%%%%%%%%%%%%%%%%%

\clearpage

\onecolumn

\appendix

\section{More Experimental Details}

\subsection{Data Format}
\label{app:data_format}

Our LOB data includes {\em order placements} and {\em transactions}, where the former one involves the orders that are submitted, while the latter one involves the orders that are eventually executed.
Common features are as follows: 
\begin{itemize}
    \item Code: Standard identifier for financial instrument. 
    \item Wind\_code: Identifier used in the Wind Information database. 
    \item Name: Asset name. 
    \item Date: The date when an order/transaction is placed/executed. 
    \item Time: Timestamp when an order/transaction is placed/executed. 
    \item Channel: The year of an order/transaction. 
\end{itemize}
The additional features of \textit{order placements} are:
\begin{itemize}
    \item Order kind: The type of order being made, e.g., limit order. 
    \item Order price: The bid/ask price of an order.  
    \item Order volume: The size of an order. 
\end{itemize}
The additional features of \textit{transactions} are:
\begin{itemize}
    \item Trade price: Price of a transaction.
    \item Trade volume: Volume of a transaction. 
    \item Ask/Bid order: Price of ask/bid order willing to trade. 
    \item Bs\_flag: Indicator of ask/bid. 
\end{itemize}

\subsection{Training Details}
\label{app:train_details}

\paragraph{Hyper-parameters} All are MLP with three layers, 256 neurons each. We use five-fold cross-validation, while 30\% of the data is left for testing. The Adam optimizer is utilized with a learning rate of $1\times 10^{-3}$. Each epoch incorporates 5 iterations. The VAE backbone is trained for 30 epochs, while the fine-tuning process of the adaptor only takes 6 epochs.

\paragraph{Hardware device} All experiments are done on Linux servers with intel-i7 CPUs and NVDIA 4090 GPUs.

\paragraph{ABIDES setup} We simulate a scale of 500 agents using the framework of ABIDES~\cite{byrd2019abides}, where the account of each agent is initially allocated a uniformly sampled amount of money equivalent to 100-500 shares.

\begin{figure}[ht]
  \centering
  \includegraphics[width=65mm]{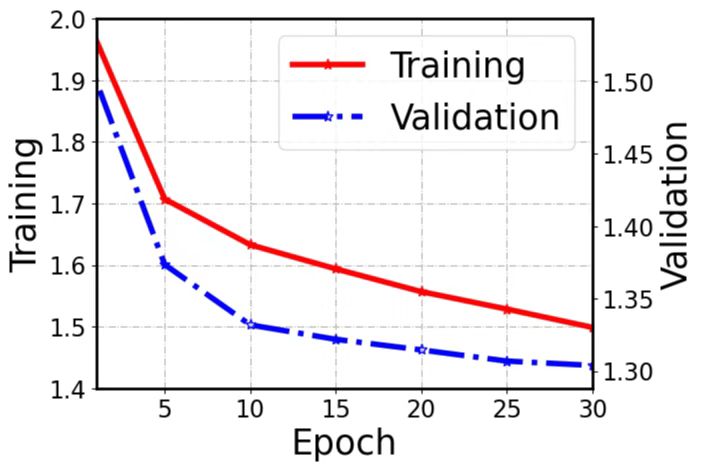} % appendix
  \includegraphics[width=65mm]{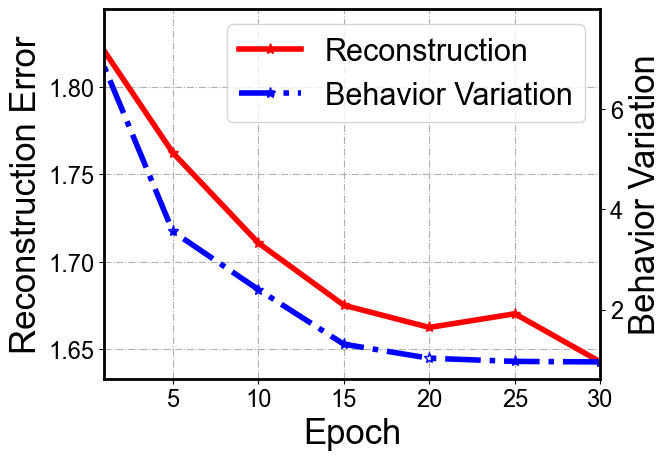} % appendix
  \caption{Some statistics during the training of the VAE backbone.}
  \label{app:fig:train_log}
  \Description{Some statistics during the training of the VAE backbone}
\end{figure}

\paragraph{Training of the VAE backbone} We illustrate some statistics during the training of the VAE backbone, in Figure~\ref{app:fig:train_log}. The left subfigure shows the training and validation loss, while the right subfigure shows the error between the reconstructed order flow and the ground truth, along with the so-called \textit{behavioral variation}, which is defined as the average absolute distance of the agent parameters between every two consecutive trading days.
It is interesting that while the quality of simulation improves, the consistency of agents' behavior patterns also tends to increase.

\paragraph{Training of the surrogate model} According to our experiments, given a set of parameters the agent model, the simulated market dynamics present strong characteristics reflected by the aforementioned stylized facts, render the training of this surrogate model very efficient. We first uniformly sampled a set of 150 groups of agent parameters (hence 150 simulations) and then trained the surrogate model for 100 epochs, 10 iterations each. The ranges of sampling for each agent parameter are as follows, respectively
\[
\alpha_F \sim \texttt{Unif}[0, 0.47],\ \alpha_C \sim \texttt{Unif}[0, 0.47],\ \tau \sim \texttt{Unif}[1, 20] (minutes),\ \sigma_N \sim \texttt{Unif}[0, 0.2],\ \beta_r \sim \texttt{Unif}[0, 0.5].
\]

\clearpage
\section{Additional Simulated Results}
\label{app:stylized}

In this section, we present some additional simulation results, in order to demonstrate the quality of the calibrated simulator.

\begin{figure}[ht]
  \centering
  \includegraphics[height=45mm]{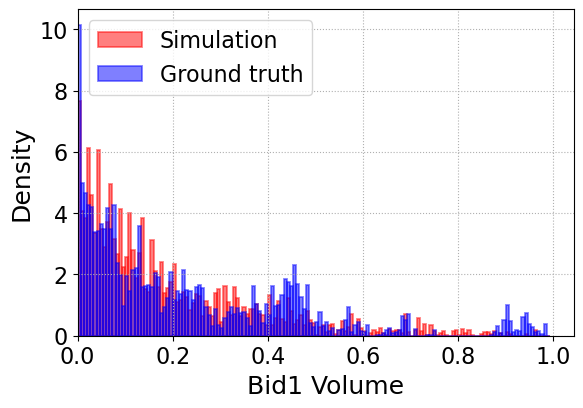} % appendix
  \includegraphics[height=45mm]{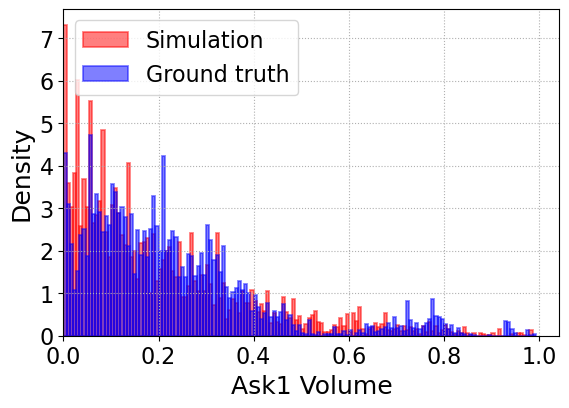} % appendix
  \caption{Distribution of volume at best bid (left) and best ask (right).}
  \label{app:fig:bestbidask}
  \Description{Distribution of volume at best bid (left) and best ask (right).}
\end{figure}

\begin{figure}[ht]
  \centering
  \includegraphics[width=60mm]{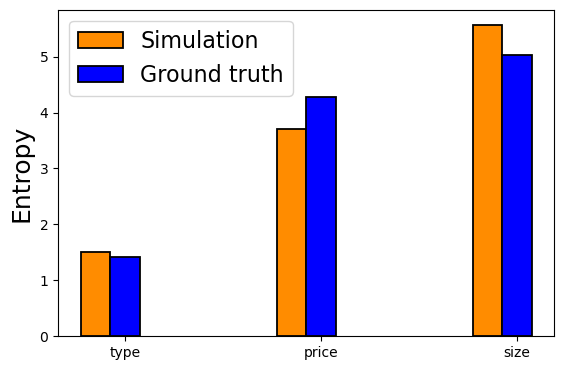} % appendix
  \caption{Entropy of order type, price, and size.
%  Order type: ask (B), sell (S), and cancel (C). 
  }
  \label{app:fig:entropy}
  \Description{Entropy of order type, price, and size.}
\end{figure}

As shown in Figure~\ref{app:fig:bestbidask}, the simulated outcome of the volumes (depths) for the best bid/ask highly resembles the ground truth.
We also calculate the entropy of the distributions collected from order types, prices, and sizes, and present the results in Figure~\ref{app:fig:entropy}.

Additionally, we also visualize the distribution of order sizes, which roughly follows the power-law distribution with the exponent chosen as 2, which is the empirical value suggested in~\cite{vyetrenko2020get}.

\begin{figure}[ht]
  \centering
  \includegraphics[width=60mm]{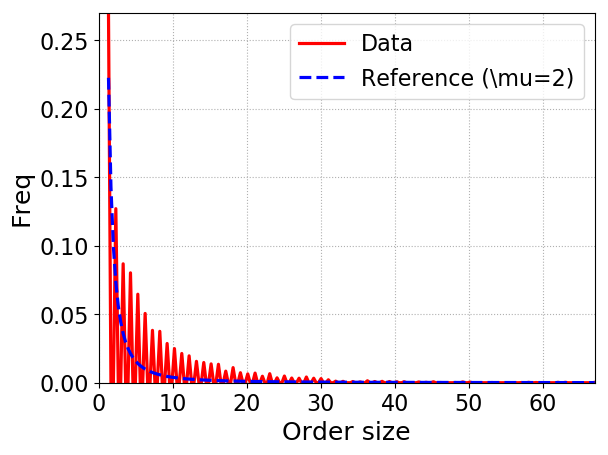} % appendix
  \caption{Order size distribution versus power-law distribution (with $\mu=2$).}
  \label{app:fig:power-law}
  \Description{Order size distribution versus power-law distribution (with $\mu=2$).}
\end{figure}

\end{document}